\documentclass{emulateapj}
\slugcomment{{\sc Accepted to ApJ:} March 25, 2014}
\usepackage{natbib}
\usepackage{amsmath}
\usepackage{graphicx}
\usepackage{apjfonts}
\usepackage{cleveref}
\newcommand{\bvri}{\protect\hbox{$BV\!RI$} }
\newcommand\kms{\ {\rm km \ s^{-1}}}
\newcommand{\about}{$\sim\!\!$~}

\renewcommand{\apjl}{ApJL}

\begin{document}
\title{HUBBLE SPACE TELESCOPE AND GROUND-BASED OBSERVATIONS 
OF \\ THE TYPE I\lowercase{ax} SUPERNOVAE SN~2005\lowercase{hk} AND SN~2008A}

\author{
Curtis~McCully\altaffilmark{1},
Saurabh~W.~Jha\altaffilmark{1},
Ryan~J.~Foley\altaffilmark{2,3},
Ryan~Chornock\altaffilmark{4},
Jon~A.~Holtzman\altaffilmark{5},
David~D.~Balam\altaffilmark{6}
David~Branch\altaffilmark{7},
Alexei~V.~Filippenko\altaffilmark{8},
Joshua~Frieman\altaffilmark{9,10},
Johan~Fynbo\altaffilmark{11},
Lluis~Galbany\altaffilmark{12,13},
Mohan~Ganeshalingam\altaffilmark{8,14}, 
Peter~M.~Garnavich\altaffilmark{15},
Melissa~L.~Graham\altaffilmark{16,17,18},
Eric~Y.~Hsiao\altaffilmark{18},
Giorgos~Leloudas\altaffilmark{11,19},
Douglas~C.~Leonard\altaffilmark{20},
Weidong~Li\altaffilmark{8,21}, 
Adam~G.~Riess\altaffilmark{22},
Masao~Sako\altaffilmark{23},
Donald~P.~Schneider\altaffilmark{24},
Jeffrey~M.~Silverman\altaffilmark{8,25,26},
Jesper~Sollerman\altaffilmark{11,27},
Thea~N.~Steele\altaffilmark{8},
Rollin~C.~Thomas\altaffilmark{28},
J.~Craig~Wheeler\altaffilmark{25}, and
Chen~Zheng\altaffilmark{29,30}\\
}

\altaffiltext{1}{Department of Physics and Astronomy, Rutgers, the
  State University of New Jersey, 136 Frelinghuysen Road, Piscataway,
  NJ 08854, USA. \\ E-mail comments or inquiries to
  cmccully@physics.rutgers.edu.}

\altaffiltext{2}{Astronomy Department, University of Illinois at
  Urbana-Champaign, 1002 West Green Street, Urbana, IL 61801 USA}

\altaffiltext{3}{Department of Physics, University of Illinois
  Urbana-Champaign, 1110 West Green Street, Urbana, IL 61801 USA}

\altaffiltext{4}{Harvard-Smithsonian Center for Astrophysics, 60
  Garden Street, Cambridge, MA 02138, USA}

\altaffiltext{5}{Department of Astronomy, MSC 4500, New Mexico State
  University, P.O. Box 30001, Las Cruces, NM 88003, USA}

\altaffiltext{6}{Dominion Astrophysical Observatory, Herzberg
  Institute of Astrophysics, 5071 West Saanich Road, Victoria, BC, V9E
  2E7, Canada}

\altaffiltext{7}{Homer L. Dodge Department of Physics and Astronomy,
  University of Oklahoma, Norman, OK 73019, USA}

\altaffiltext{8}{Department of Astronomy, University of California,
  Berkeley, CA 94720-3411, USA}

\altaffiltext{9}{Kavli Institute for Cosmological Physics and
  Department of Astronomy and Astrophysics, University of Chicago,
  5640 South Ellis Avenue, Chicago, IL 60637, USA}

\altaffiltext{10}{Center for Particle Astrophysics, Fermi National
  Accelerator Laboratory, P.O. Box 500, Batavia, IL 60510, USA}

\altaffiltext{11}{Dark Cosmology Centre, Niels Bohr Institute,
  University of Copenhagen, Juliane Maries Vej 30, DK-2100 Copenhagen
  \O, Denmark}

\altaffiltext{12}{Institut de F\'isica d'Altes Energies, Universitat
  Aut\`onoma de Barcelona, E-08193 Bellaterra (Barcelona), Spain}

\altaffiltext{13}{Centro Multidisciplinar de Astrof\'isica, Instituto
  Superior T\'ecnico, Av. Rovisco Pais 1, 1049-001 Lisbon, Portugal}

\altaffiltext{14}{Lawrence Berkeley National Laboratory,
  Berkeley, CA 94720, USA}

\altaffiltext{15}{Department of Physics, University of Notre Dame,
  Notre Dame, IN 46556, USA}

\altaffiltext{16}{Las Cumbres Observatory Global Telescope Network,
  Goleta, CA 93117, USA}

\altaffiltext{17}{Department of Physics, University of California,
  Santa Barbara, CA 93106, USA}

\altaffiltext{18}{Carnegie Observatories, Las Campanas Observatory,
  Colina El Pino, Casilla 601, Chile}

\altaffiltext{19}{The Oskar Klein Centre, Department of Physics,
  Stockholm University, AlbaNova, SE-10691 Stockholm, Sweden}

\altaffiltext{20}{Department of Astronomy, San Diego State University,
  San Diego, CA 92182, USA}

\altaffiltext{21}{Deceased December 12, 2011}

\altaffiltext{22}{Department of Physics and Astronomy, Johns Hopkins
  University, Baltimore, MD 21218, USA}

\altaffiltext{23}{Department of Physics and Astronomy, University of
  Pennsylvania, 209 South 33rd Street, Philadelphia, PA 19104, USA}

\altaffiltext{24}{Department of Astronomy and Astrophysics, and
  Institute for Gravitation and the Cosmos, The Pennsylvania State
  University, University Park, PA 16802, USA}

\altaffiltext{25}{Department of Astronomy,
  University of Texas at Austin, Austin, TX 78712, USA}

\altaffiltext{26}{National Science Foundation Astronomy and
  Astrophysics Postdoctoral Fellow}

\altaffiltext{27}{The Oskar Klein Centre, Department of Astronomy,
  Stockholm University, AlbaNova, SE-10691 Stockholm, Sweden}

\altaffiltext{28}{Computational Cosmology Center, Lawrence Berkeley
  National Laboratory, 1 Cyclotron Road MS50B-4206, Berkeley, CA,
  94720, USA}

\altaffiltext{29}{Kavli Institute for Particle Astrophysics and
  Cosmology, SLAC National Accelerator Laboratory, 2575 Sand Hill
  Road, Menlo Park, CA 94025, USA}

\altaffiltext{30}{Department of Physics, Stanford University,
  Stanford, CA 94305, USA}

\begin{abstract}
We present {\it Hubble Space Telescope} ({\it HST}) and ground-based
optical and near-infrared observations of SN~2005hk and SN~2008A,
typical members of the Type Iax class of supernovae (SNe). Here we focus on late-time observations, where these
objects deviate most dramatically from all other SN types. Instead of the
dominant nebular emission lines that are observed in other SNe at
late phases, spectra of
SNe~2005hk and 2008A show lines of \ion{Fe}{2}, \ion{Ca}{2}, and
\ion{Fe}{1} more than a year past maximum light, along with
narrow [\ion{Fe}{2}] and [\ion{Ca}{2}] emission. We use spectral
features to constrain the temperature and density of the ejecta, and
find high densities at late times, with $n_e \gtrsim
10^{9}$~cm$^{-3}$.  Such high densities should yield enhanced cooling
of the ejecta, making these objects good candidates to observe the
expected ``infrared catastrophe,'' a generic feature of SN~Ia models. However,
our {\it HST} photometry of SN~2008A does not match the predictions of
an infrared catastrophe. Moreover, our {\it HST} observations rule out a
``complete deflagration'' that fully disrupts the white dwarf for these peculiar SNe, showing no
evidence for unburned material at late times. 
Deflagration explosion models that leave
behind a bound remnant can match some of the observed properties of
SNe~Iax, but no published model is consistent with all of our
observations of SNe~2005hk and 2008A.

\end{abstract}

\section{Introduction}
\label{sec:introduction}

The use of Type Ia supernovae (SNe~Ia) as distance indicators has
revolutionized cosmology with the discovery that the expansion of the
universe is currently accelerating, probably driven by dark energy
\citep{Riess98,Perlmutter99}. Sufficiently large samples have now
been collected such that systematic uncertainties are beginning to
dominate the statistical uncertainties in SN~Ia distances
\citep[e.g.,][]{Wood-Vasey07,Kessler09,Conley11}.  Perhaps one of the
most fundamental systematic uncertainties stems from the lack of
detailed understanding of SN~Ia progenitor systems and explosion
mechanism. Though exploding white dwarfs typically produce normal SNe
Ia (by which we include all objects that fall on the one-parameter
family correlating luminosity with light-curve width;
\citealt{Phillips93}), we are amassing growing evidence that other SNe
are also consistent with a white dwarf origin. Understanding what
makes these thermonuclear explosions different can shed light on both
normal SNe~Ia and more general outcomes of stellar evolution.

SN~2002cx was labeled ``the most peculiar known SN~Ia'' by
\citet{Li03}; see also
  \citet{Filippenko03}.  While SN~2002cx was peculiar, it is not
  unique. SN~2002cx is the prototype for the largest class of peculiar
  SNe, which we have dubbed ``Type Iax'' supernovae \citep[for a full
    description of this class, see][]{Foley13}. These are weak
  explosions with luminosities that can fall more than a magnitude
  below the Phillips relation for normal SNe~Ia with similar decline
  rate, and they have ejecta velocities roughly half those of normal
  SNe~Ia \citep{Jha06}. Still, near peak brightness, SNe~Iax are
  similar to SNe~Ia in the general characteristics of their light
  curves and spectral features. However, the late-time properties of
  SNe~Iax are unmatched by any other previously discovered SN
  class. Instead of entering a nebular phase dominated by broad
  forbidden lines of iron-peak elements, the spectrum of SN~2002cx at
  \about 250~days past maximum brightness\footnote{Throughout this
    paper, SN phases are given in rest-frame days past $B$-band
    maximum light.} was dominated by \emph{permitted} \ion{Fe}{2},
  with very low expansion velocities \about 700$\kms$, much lower than
  ever observed in normal SNe~Ia \citep{Jha06}. In addition, the
  late-time spectrum of SN~2002cx showed hints of low-velocity
  \ion{O}{1}, also unprecedented in SNe~Ia, and perhaps an indication
  of unburned material in the inner regions of the white dwarf.

A variety of models have been proposed to explain the origins of
SNe~Iax. \citet{Branch04} and \citet{Jha06} suggested that these
objects might be explained by pure deflagration models. These models
do not explain normal SNe~Ia well: a pure deflagration model typically
produces much less nickel than is required for the luminosity of a
normal SN~Ia \citep{Gamezo04}. Moreover, the highly turbulent and
convoluted thermonuclear burning front in these models yields clumpy,
well-mixed ejecta, with unburned material, partially burned material,
and fully burned (to the iron peak) material at all layers
\citep{Roepke08}, and this mixing is not observed in normal SNe~Ia
\citep{Gamezo04}. One of the strongest constraints for a
pure-deflagration model is the prediction of unburned material (carbon
and oxygen) in the innermost layers, which should be easily detectable
in late-time spectra and yet has never been observed in normal SNe~Ia
\citep{Gamezo04}.

The problems with the pure deflagration model for normal SNe~Ia may
become strengths for SNe~Iax \citep{Jha06}. The low production of
nickel and the low luminosity are key traits of SNe~Iax. Large amounts
of mixing of partially burned, fully burned, and possibly unburned
material are observed in all layers of the ejecta. The clumpiness
predicted by the pure deflagration model could explain the high
densities seen at late times \citep{Jha06,Phillips07}. In SN~2002cx,
there was a tentative detection of \ion{O}{1} $\lambda$7774
\citep{Jha06} and a hint of the line in SN~2005hk, a prototypical
SN~Iax \citep{Phillips07, Stanishev07, Sahu08}. However, ``complete
deflagration'' models that fully unbind the white dwarf do not predict the
high densities at late times seen in SNe~Iax, and therefore suggest
that oxygen in the inner layers should be revealed by [\ion{O}{1}]
$\lambda\lambda$6300, 6363 emission \citep{Kozma05}.

More recently, \citet{Jordan12} find that if a detonation is not
triggered, the explosion is often not powerful enough to unbind the
star, leading to the low luminosities and ejecta velocities like those
found in SNe~Iax. \citet{Kromer13} also study the three-dimensional deflagration 
of a Chandrasekhar-mass white dwarf. Using radiative transfer models,
they find that they can reproduce the luminosity, the early-time
light curve, and the early-time spectra of SN~2005hk. Similarly, one
of the key features of their explosion simulation is a bound remnant.

The discovery of SN~2008ha sparked controversy about the nature of
these peculiar SNe. SN~2008ha was spectroscopically a SN~Iax, but it
was the most extreme member to date \citep{Foley09, Valenti09}, with
maximum-light expansion velocities of just \about 2000$\kms$, less
than half that of even typical SNe~Iax. In addition, SN~2008ha was 3
mag fainter than SN~2002cx, with a much more rapid light-curve
decline rate. Based on the energetics, and the spectral similarity of
SN~2008ha to SN 1997D at late times, \citet{Valenti09} argued that
SN~2008ha was actually a core-collapse SN, rather than a thermonuclear
one. Indeed, \citet{Moriya10} were able to recreate the kinetic energy
of SN~2008ha in a core-collapse simulation with large amounts of
fallback onto a newly formed black hole \citep{Foley09,Valenti09}. If
this model holds for SN~2008ha, \citet{Valenti09} argued by extension
that all SNe~Iax might actually be core-collapse SNe. Further support
for this idea comes from the fact that, like SN~2008ha, SNe~Iax are
found almost exclusively in late-type galaxies, similar to
core-collapse SNe \citep{Jha06, Foley09, Foley13}. Using H$\alpha$ maps of a sample of host galaxies of SNe Iax, \citet{Lyman13} find a statistical association with star forming regions similar to that of SNe IIP. The two objects we focus on in this work, SN 2005hk and SN 2008A, are both in star-forming galaxies, but there is no evidence for star formation at the location of either object \citep{Lyman13}.

\begin{table*}[!t]
\centering
\small
\caption{Ground-based Optical Photometry of SN~2005hk from the SDSS-II
  SN Survey \label{tab-phot05hk}}
\begin{tabular}{ c  c  c  c  c  c c c }
\hline
\hline
  Date  &  MJD     &  Phase   &  $u$    &  $g$    &  $r$    &  $i$    &  $z$ \\
  (UT)  &  (days)  &  (days)  &  (mag)  &  (mag)  &  (mag)  &  (mag)  &  (mag)\\
\hline
2005 Oct 28 & 53671.34 & \phs$-$12    & 18.586(0.036) & 18.733(0.012) & 18.954(0.019) & 19.284(0.027) & 19.604(0.104) \\
2005 Oct 31 & 53674.24 & \phs\phs$-$9 & 17.018(0.031) & 16.977(0.012) & 17.097(0.006) & 17.348(0.009) & 17.570(0.018) \\
2005 Nov 02 & 53676.33 & \phs\phs$-$7 & 16.649(0.031) & 16.511(0.004) & 16.592(0.004) & 16.807(0.005) & 17.010(0.011) \\
2005 Nov 05 & 53679.30 & \phs\phs$-$4 & 16.409(0.031) & 16.046(0.009) & 16.148(0.009) & 16.393(0.015) & 16.570(0.014) \\
2005 Nov 07 & 53681.29 & \phs\phs$-$2 & 16.400(0.031) & 15.903(0.004) & 15.977(0.003) & 16.214(0.005) & 16.343(0.009) \\
2005 Nov 11 & 53685.25 & \phs\phs$+$1 & 16.524(0.031) & 15.777(0.018) & 15.743(0.015) & 16.003(0.015) & 16.120(0.010) \\
2005 Nov 23 & 53697.25 & \phs$+$12    & 18.423(0.031) & 16.778(0.015) & 15.810(0.006) & 15.820(0.006) & 15.917(0.016) \\
2005 Nov 26 & 53700.25 & \phs$+$15    & 18.991(0.038) & 17.125(0.022) & 15.988(0.013) & 15.888(0.014) & 15.967(0.018) \\
2005 Dec 01 & 53705.23 & \phs$+$20    & 19.637(0.050) & 17.625(0.013) & 16.307(0.003) & 16.170(0.008) & 16.190(0.009) \\
\\
2006 Aug 28 & 53975.32 & $+$287       &    \nodata    & 21.812(0.069) & 20.840(0.042) & 20.375(0.039) & 20.327(0.149) \\
2006 Sep 12 & 53990.34 & $+$302       &    \nodata    & 22.051(0.218) & 21.194(0.097) & 20.506(0.064) & 20.826(0.270) \\
2006 Sep 16 & 53994.35 & $+$306       &    \nodata    & 22.242(0.103) & 21.121(0.062) & 20.528(0.044) & 20.666(0.192) \\
2006 Sep 18 & 53996.33 & $+$308       &    \nodata    & 22.260(0.127) & 21.205(0.064) & 20.650(0.062) & 20.900(0.301) \\
2006 Sep 20 & 53998.30 & $+$310       &    \nodata    & 22.370(0.113) & 21.206(0.051) & 20.660(0.049) & 20.873(0.211) \\
2006 Sep 27 & 54005.33 & $+$317       &    \nodata    & 22.299(0.093) & 21.251(0.049) & 20.674(0.042) & 20.695(0.162) \\
2006 Sep 30 & 54008.29 & $+$320       &    \nodata    & 22.439(0.120) & 21.198(0.066) & 20.806(0.057) & 20.856(0.220) \\
2006 Oct 02 & 54010.28 & $+$322       &    \nodata    & 22.441(0.184) & 21.225(0.065) & 20.723(0.055) & 20.615(0.151) \\
2006 Oct 04 & 54012.28 & $+$324       &    \nodata    & 22.270(0.315) & 21.321(0.157) & 20.677(0.070) & 20.680(0.207) \\
2006 Oct 12 & 54020.28 & $+$332       &    \nodata    & 22.646(0.288) & 21.371(0.100) & 20.800(0.077) & 20.974(0.318) \\
2006 Oct 16 & 54024.36 & $+$336       &    \nodata    &    \nodata    & 21.360(0.098) & 21.000(0.092) &    \nodata    \\
2006 Oct 22 & 54030.28 & $+$342       &    \nodata    & 22.533(0.112) & 21.466(0.063) & 21.017(0.057) & 21.230(0.269) \\
\hline
\end{tabular}
\tablecomments{1$\sigma$ photometric uncertainties are given in parentheses.}
\end{table*}

\citet{Foley10} published a new set of spectra from earlier epochs of
SN~2008ha showing strong evidence for both \ion{Si}{2} and \ion{S}{2}
at maximum light. While some core-collapse SNe do show weak
\ion{Si}{2} lines, the sulfur lines are usually considered hallmarks
of thermonuclear burning in a C/O white dwarf \citep[Sulfur may also be present in the ejecta of other type of SNe, but has never been clearly detected in other type of SNe][]{Foley10}; these
lines were also seen in SN~2007qd, another SN~Iax very similar to
SN~2008ha \citep{McClelland10}.  \citet{Foley10} proposed that SN~2008ha
is better explained by a failed deflagration than a core-collapse
model. The host-galaxy distribution is also similar to that of some SNe~Ia,
specifically SN 1991T-like objects \citep{Foley09}, and the
SN~Iax~2008ge exploded in an S0 galaxy with no sign of local star
formation to deep limits, inconsistent with a massive star origin
\citep{Foley10_ge}.

Because these SNe deviate most dramatically from normal SNe (both
core-collapse and thermonuclear) at late times, here we present late-time 
observations of SNe~2005hk and 2008A \citep{Foley13}, both
typical SNe~Iax very similar to SN~2002cx, to constrain the nature of
this class of SNe. We look for evidence of the [\ion{O}{1}]
$\lambda6300$ line that is predicted for a complete pure deflagration
(and is usually seen in core-collapse SNe). We also use late-time
spectra to infer the composition, velocity structure, density,
and temperature of the ejecta.

Besides providing insight to the progenitor systems and explosion
mechanisms of white dwarf SNe, SNe~2005hk and 2008A may be ideal
candidates to observe the ``infrared (IR) catastrophe'' predicted by
\citet{Axelrod80}, a thermal instability that changes the dominant
cooling mechanism from optical lines to far-IR fine-structure
lines such as [\ion{Fe}{1}] 24~$\mu$m and [\ion{Fe}{2}] 26~$\mu$m
\citep{Sollerman04}. This phenomenon has never been observed in normal
SNe~Ia even out to 700~days past maximum \citep{Leloudas09}. Because
SNe~Iax remain at high densities at late times, the objects in this
class should cool faster than normal SNe~Ia and should undergo this
instability sooner. Using two epochs of late-time {\it HST}
observations, we compare the color evolution of SN~2008A to the
predictions of IR-catastrophe models.

Throughout this paper we adopt $H_0 = 73 \kms$ Mpc$^{-1}$ and correct
redshifts ($z$) using the Virgo+GA infall model of \citet{Mould00} via
NED\footnote{http://ned.ipac.caltech.edu.} to estimate distances to
the SN host galaxies.

\section{Observations and Data Reduction}

\subsection{Ground-based Optical Photometry and Spectroscopy}

For SN~2005hk, we supplement the ground-based optical photometry of
\citet{Phillips07}, \citet{Stanishev07}, and \citet{Sahu08} with
observations of the equatorial ``Stripe 82'' from the SDSS-II SN
survey \citep{Frieman08, Sako08}, following the reduction procedure
detailed by \citet{Holtzman08}. The data comprise $ugriz$ photometry
\citep{Fukugita96} from the 2005 and 2006 SDSS-II observing seasons,
and are presented in Table~\ref{tab-phot05hk}.

The ground-based early-time \bvri observations of SN~2008A were
obtained with the 0.76~m Katzman Automatic Imaging Telescope
\citep{Filippenko01} and 1~m Nickel telescope at Lick Observatory
\citep{Ganeshalingam10}. $BV\!ri$~observations of SN~2008A were also
taken as part of the CfA4 survey \citep{Hicken12}.

\begin{figure*}[!h]
\centering
\includegraphics[width=0.78\textwidth]{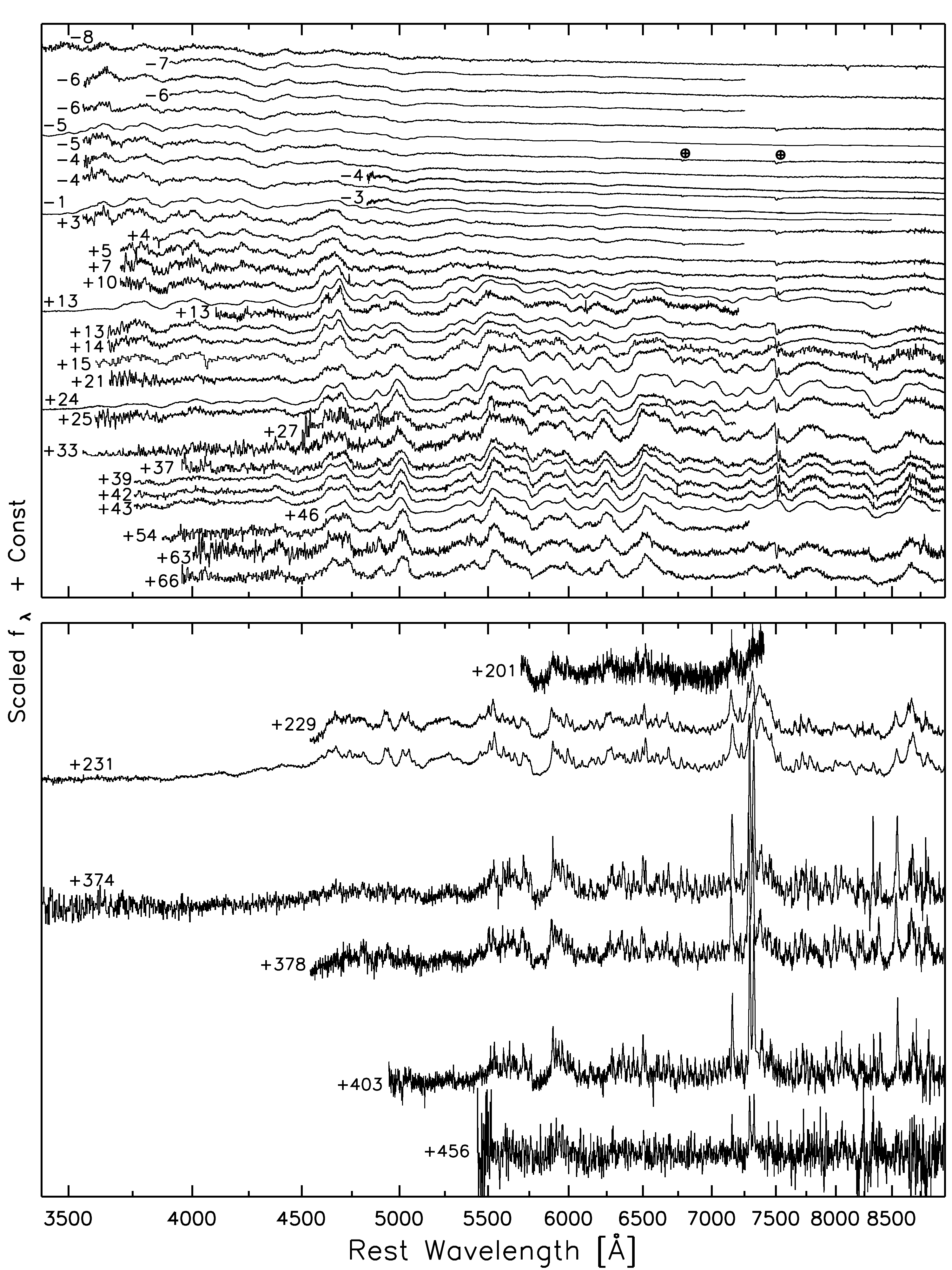}
\caption{Spectral time series of SN~2005hk, labeled by phase (in
  days). Both the phases and wavelengths are corrected to the SN rest
  frame using $z = 0.01176$ from the [\ion{Ca}{2}] $\lambda\lambda$
  7291, 7323 lines. We include spectra from \citet{Chornock06},
  \citet{Phillips07}, \citet{Sahu08}, and \citet{Silverman12}. The top
  panel shows the early-time spectra and the bottom panel shows the
  late-time spectra.
}
\label{fig:spec_05hk}
\end{figure*}

\begin{figure*}[!h]
\centering
\includegraphics[width = 0.70\textwidth]{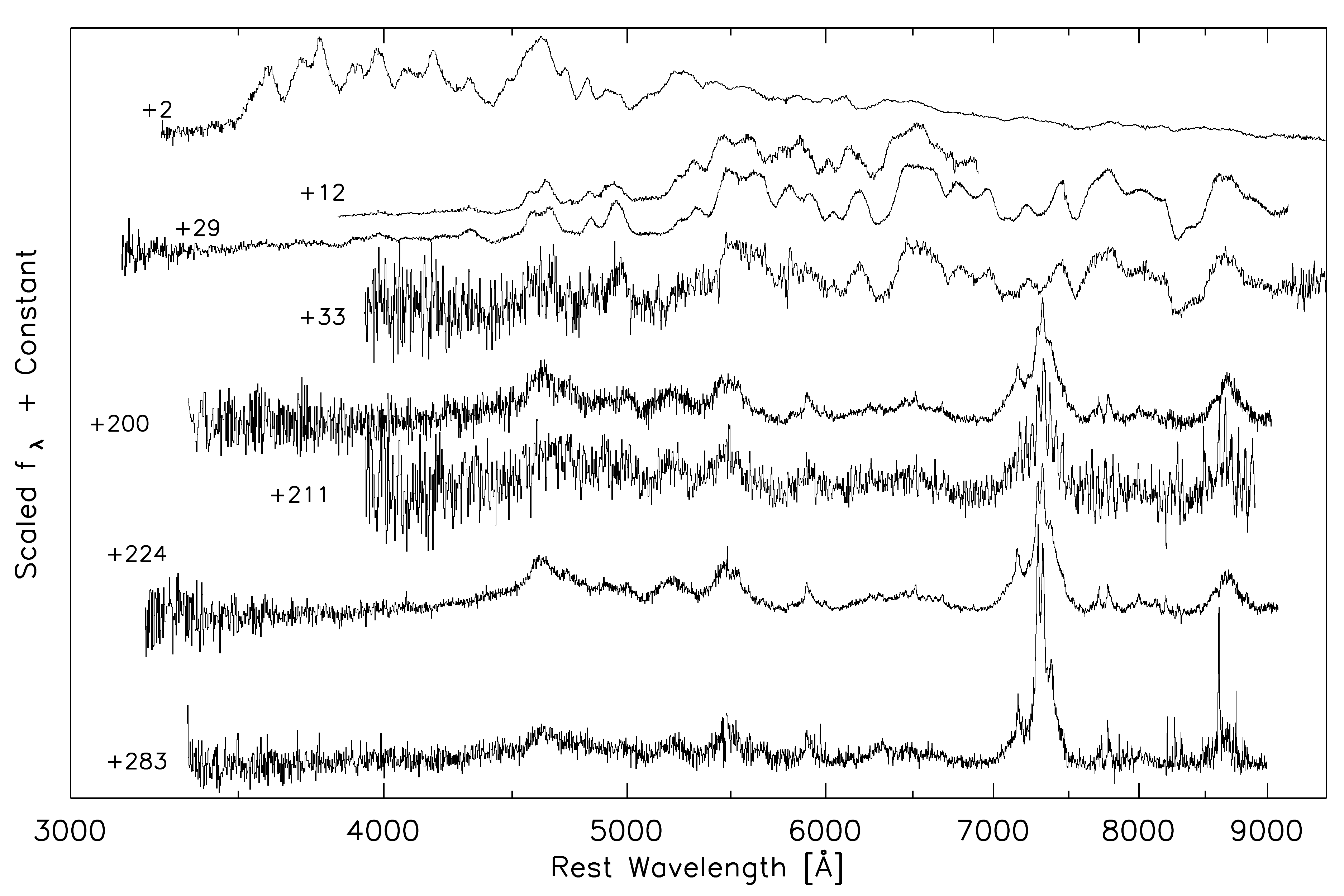}
\caption{Spectral time series of SN~2008A, labeled by phase (in days).
  Both the phases and wavelengths are corrected to the SN rest frame
  using $z = 0.01825$ determined
  from the [\ion{Ca}{2}] $\lambda\lambda$ 7291, 7323
  lines.
}
\label{fig:spec_08a}
\end{figure*}

Optical spectra of SN~2005hk were obtained with the Low Resolution 
Imaging Spectrometer \citep[LRIS;][]{Oke95} (sometimes in polarimetry
mode, LRISp) on the Keck I 10~m telescope and the Deep Imaging 
Multi-Object Spectrograph \citep[DEIMOS;][]{Faber03} on the Keck II
10~m telescope. Spectra of SN~2008A were obtained with the Kast
double spectrograph \citep{Miller93} on the Lick 3~m Shane telescope, 
Keck I (+ LRIS), and the Dominion Astrophysical Observatory (DAO) 
Plaskett 1.8~m telescope. Data-reduction procedures for the Lick and Keck
spectra are presented by \citet{Silverman12}; the DAO spectroscopy was
reduced with standard techniques. Logs of the observations are
provided in Tables~\ref{tab:spec05hk} and \ref{tab:spec08a}, and
spectral time series are shown in Figures~\ref{fig:spec_05hk}
and \ref{fig:spec_08a}. The former figure includes supplementary
spectroscopy of SN~2005hk from \citet{Chornock06}, \citet{Phillips07},
\citet{Sahu08}, and \citet{Maund10}. The listed phases are in the SN
rest frame, referenced to $B$ maximum light, which occurred on MJD
53684.2 for SN~2005hk and MJD 54478.3 for SN~2008A.

\begin{table*}[!h]
\centering
\caption{Late-time Spectroscopic Observations of SN~2005hk \label{tab:spec05hk}}
\begin{tabular}{c c c c c c c}
\hline
UT & MJD & Phase & Telescope/Instrument & Exposure & Range & Resolution\\
 &  & (days) &  & (s) & (\AA) & (\AA)\\
\hline
\hline
2006 Jun 01 & 53887.62 & $+$201 & Keck I/LRISp & \phs800 & 5765--7492 & 3 \\
2006 Jul 01 & 53917.58 & $+$231 & Keck I/LRISp & 4200 & 3260--9276 & 7 \\
2006 Nov 23 & 54062.22 & $+$374 & Keck I/LRIS & 1800 & 3150--9250 & 7\\
2006 Dec 23 & 54092.21 & $+$403 & Keck II/DEIMOS & 1800 & 5000--9300 & 3\\
2007 Feb 14 & 54145.23 & $+$456 & Keck I/LRIS & 1800 & 5500--9240 & 7\\
\hline
\end{tabular}
\end{table*}

\begin{table*}[!h]
\centering
\caption{Spectroscopic Observations of SN~2008A \label{tab:spec08a}}
\begin{tabular}{c c c c c c c}
\hline
UT & MJD & Phase & Telescope/Instrument & Exposure & Range & Resolution\\
 &  & (days) &  & (s) & (\AA) & (\AA)\\
\hline
\hline
2008 Jan 15 & 54480.28 & \phs\phs+2 & Lick/Kast & 1800+1350 & 3320--10500 & 6--12\\
2008 Jan 25 & 54490.11 & \phs+12 & Plaskett/Cassegrain & 9600 & 3900--7025 & 6\\
2008 Feb 12 & 54508.27 & \phs+29 & Keck I/LRIS & 200 & 3075--9340 & 7\\
2008 Feb 16 & 54512.23 & \phs+33 & Lick/Kast & 1800 & 3300--10500 & 6--12\\
2008 Aug 03 & 54681.63 & +200 & Keck I/LRIS & 600+300 & 3270--9196 & 7\\
2008 Aug 15 & 54693.18 & +211 & NOT/ALFOSC & 3600 & 4000-9050 & 8\\
2008 Aug 28 & 54706.51 & +224 & Keck I/LRIS & 1200 & 3270--9196 & 7\\
2008 Oct 27 & 54766.28 & +283 & Keck I/LRIS & 1800 & 3100--9160 & 7\\
\hline
\end{tabular}
\end{table*}

\begin{table*}[!h]
\centering
\caption{Late-time {\it HST} Observations of SN~2005hk and
  SN~2008A\label{tab:hstphot}}
\begin{tabular}{c c c c c c c}
\hline
Object & UT & MJD & Phase  & Instrument/Filter & Exposure & Magnitude\\
       &    &     & (days) &                   &    (s)   & (mag)\\
\hline
\hline
SN~2005hk & 2007 May 31 & 54251.11 & $+$560 & WFPC2/F450W & 1600  &$>25.50$\\
SN~2005hk & 2007 May 31 & 54251.12 & $+$560 & WFPC2/F675W & \phs900   &$>24.80$\\
SN~2005hk & 2007 May 31 & 54251.91 & $+$561 & WFPC2/F555W & \phs920   &$>25.50$\\
SN~2005hk & 2007 May 31 & 54251.92 & $+$561 & WFPC2/F814W & 1400  &$24.91(0.22)$\\
SN~2005hk & 2007 Jun 25 & 54276.09 & $+$585 & NIC2/F110W  & 5120  &$>25.86$\\
SN~2005hk & 2007 Jun 27 & 54278.68 & $+$588 & WFPC2/F555W & \phs460   &$>25.20$\\
SN~2005hk & 2007 Jun 27 & 54278.69 & $+$588 & WFPC2/F814W & \phs700   &$>24.55$\\
SN~2005hk & 2007 Aug 13 & 54325.45 & $+$634 & WFPC2/F606W & 4800  &$26.35(0.15)$\\
SN~2005hk & 2007 Aug 13 & 54325.91 & $+$634 & WFPC2/F814W & 1600  &$>25.10$\\
SN~2005hk & 2007 Aug 15 & 54327.24 & $+$636 & NIC2/F160W  & 7680  &$>25.08$\\ 
\\
SN~2008A  & 2009 Feb 19 & 54881.15 & $+$396 & WFPC2/F555W  & 1000  &$24.20(0.11)$\\
SN~2008A  & 2009 Feb 19 & 54881.17 & $+$396 & WFPC2/F791W  & \phs900   &$23.10(0.10)$\\
SN~2008A  & 2009 Feb 19 & 54881.21 & $+$396 & WFPC2/F622W  & \phs800   &$24.14(0.13)$\\
SN~2008A  & 2009 Feb 19 & 54881.24 & $+$396 & WFPC2/F850LP & 1450  &$23.40(0.26)$\\
SN~2008A  & 2009 Feb 20 & 54882.88 & $+$397 & WFPC2/F439W  & 1500  &$>24.35$\\
SN~2008A  & 2009 Aug 18 & 55061.34 & $+$573 & ACS/F625W    & 3530  &$26.15(0.06)$\\
SN~2008A  & 2009 Aug 18 & 55061.41 & $+$573 & ACS/F555W    & 3750  &$26.25(0.05)$\\
SN~2008A  & 2009 Aug 18 & 55061.54 & $+$573 & ACS/F775W    & 2484  &$26.25(0.11)$\\
SN~2008A  & 2009 Aug 18 & 55061.61 & $+$573 & WFC3IR/F110W & 8335  &$26.00(0.21)$\\
\hline
\end{tabular}
\tablecomments{1$\sigma$ photometric uncertainties are given in
  parentheses. Upper limits are 3$\sigma$.}
\end{table*}

\subsection{Late-time {\it HST} Observations}

Our {\it HST} observations of SNe~2005hk and
2008A include optical photometry from WFPC2 and Advanced Camera for Surveys (ACS)/WFC and
near-IR photometry from NICMOS and WFC3/IR, which are
presented in Table~\ref{tab:hstphot}. WFPC2 and NICMOS observations of
SN~2005hk were taken as part of {\it HST} program GO-11133 (PI:
Jha). Additional WFPC2 observations of SN~2005hk were available from
{\it HST} snapshot program GO-10877 (PI: Li). SN~2008A was observed in the
optical using WFPC2 and ACS and in the near-IR using WFC3/IR as part
of {\it HST} program GO-11590 (PI: Jha). 

\begin{figure*}[!ht]
\centering
\includegraphics[width=0.96\textwidth]{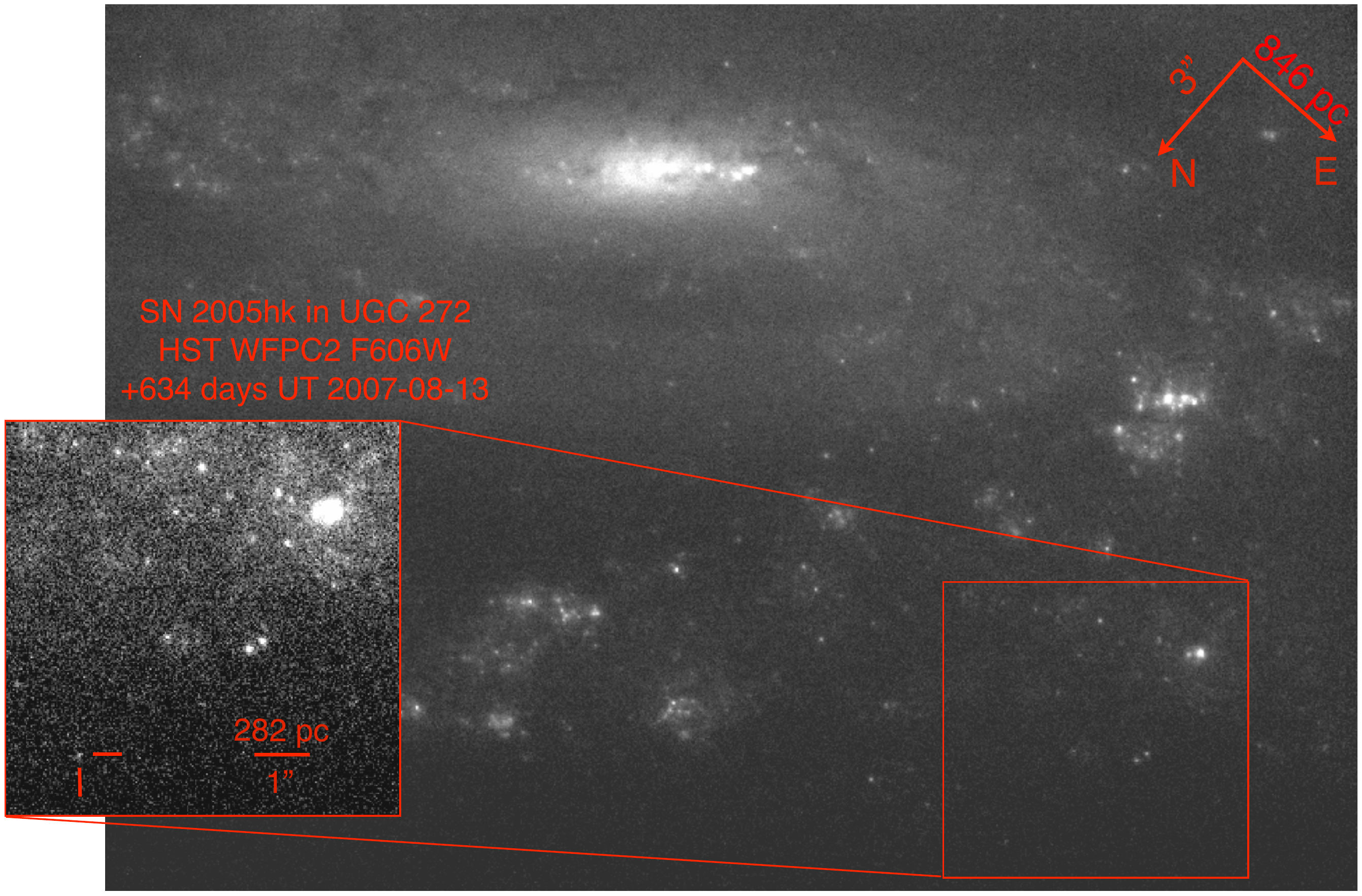}
\caption{F606W observations of SN~2005hk in UGC 272 ($z =0.013$) using
  the WFPC2 instrument on {\it HST} taken 634 days after $B$ maximum. The SN
  is marked in the inset, which has been rescaled to better show the
  SN. \label{fig:05hk_hst}}
\end{figure*}

\begin{figure*}[!ht]
\centering
\includegraphics[width=0.96\textwidth]{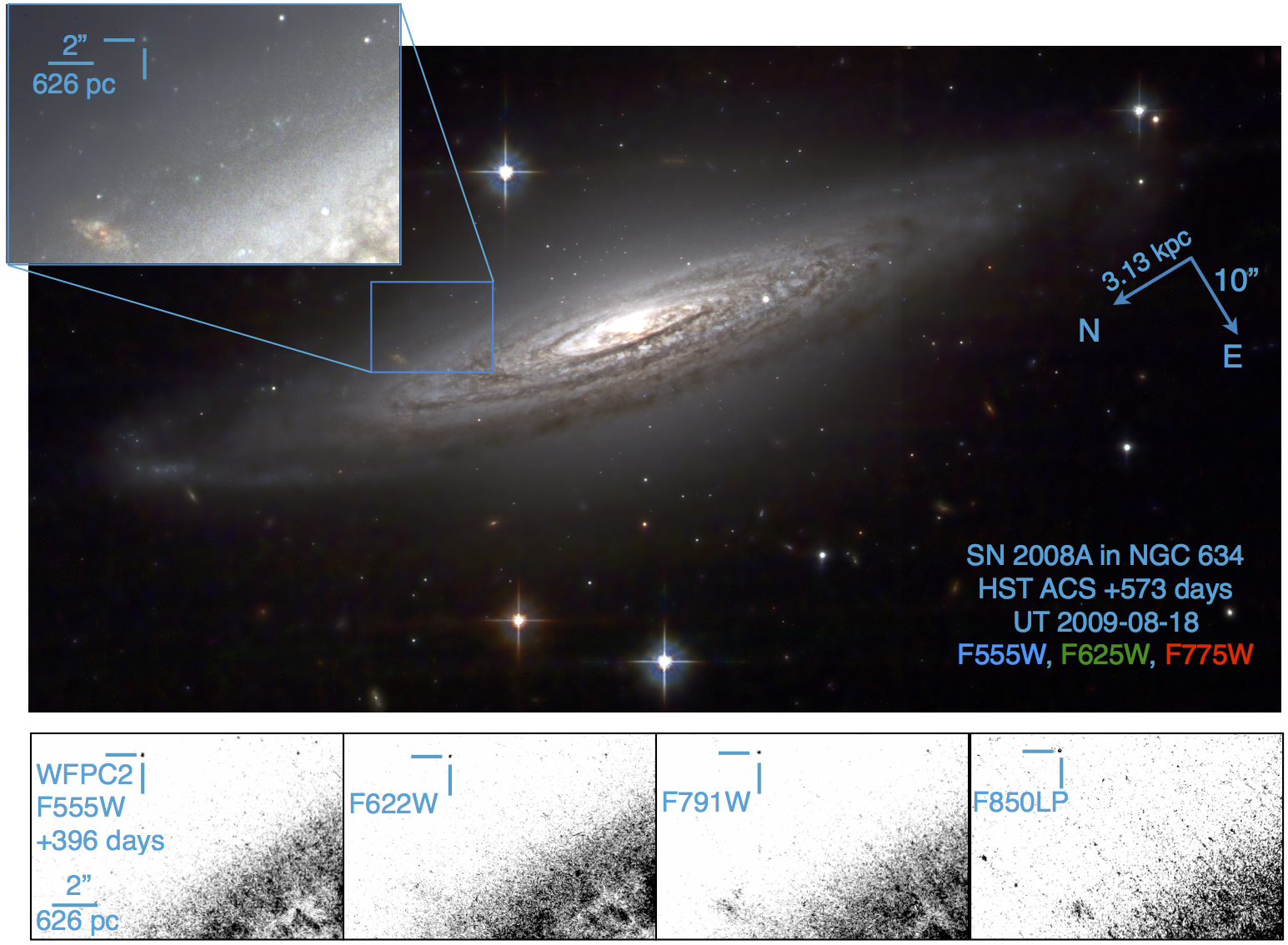}
\caption{{\it HST} observations of SN~2008A in NGC 634 ($z =
  0.016$). The top image is combined from $Vri$ (F555W, F625W, F775W)
  data taken with ACS, 573 days after maximum light. SN~2008A is
  marked in the inset image, in the outskirts of its host. Color image was produced using STIFF \citep{stiff}. An unsharp
  mask filter has been applied to the color figure to emphasize faint
  sources for display purposes. The bottom panels show WFPC2
  observations of SN~2008A taken 396 days after maximum light, in the
  F555W, F622W, F791W, and F850LP filters. Like SN~2005hk, SN~2008A is
  located in the outskirts of a spiral
  galaxy. \label{fig:08A_optical_hst}}
\end{figure*}

\begin{figure*}[!ht]
\centering
\includegraphics[width=0.85\textwidth]{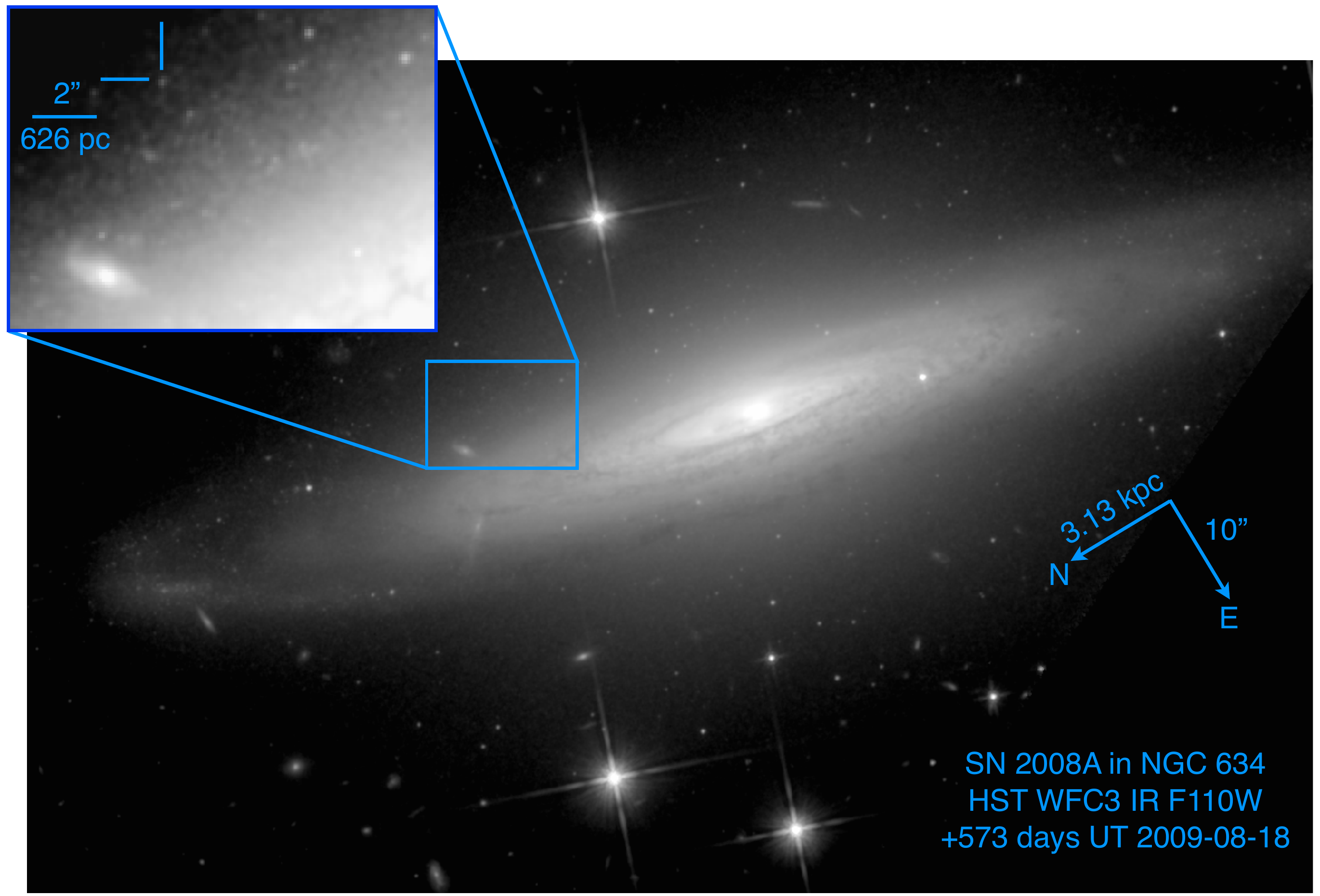}
\caption{{\it HST} near-infrared observations of SN~2008A in
  NGC 634, taken with WFC3/IR in the F110W passband, 573 days after
  maximum light. \label{fig:08A_ir_hst}}
\end{figure*}

The {\it HST} observations were combined (with cosmic-ray rejection
and subsampling) using MultiDrizzle \citep{Multidrizzle} with standard
parameters. The resulting images of SN~2005hk and SN~2008A are shown
in Figures \ref{fig:05hk_hst}-\ref{fig:08A_ir_hst}. We performed aperture photometry on all of the
images using the APPHOT task in IRAF\footnote{IRAF is distributed by
  the National Optical Astronomy Observatory, which is operated by
  the Association of Universities for Research in Astronomy, Inc.,
  under cooperative agreement with the National Science
  Foundation.}. We used aperture corrections based on encircled
energies given by \citet{Holtzman95} for WFPC2, \citet{Sirianni05} for
ACS/WFC, and the WFC3 Instrument handbook \citep{WFC3Handbook} for WFC3/IR. 
For NICMOS, we generated a point-spread function model using the TinyTim
software \citep[][but see \citealt{Hook08}]{Krist93} and then directly
measured encircled energies from the model. For the ACS and WFPC2 data
we also corrected for the charge-transfer inefficiency of the
detectors, using the models of \citet{Dolphin09} for WFPC2 and
\citet{Chiaberge09} for ACS/WFC. To derive upper limits in cases
without a significant detection of the SN, we injected fake stars with
a range of magnitudes into the images. These artificial stars were
photometered in the same way as the SNe, with the standard deviation
of the recovered magnitudes used to estimate the photometric
uncertainty. We list our derived photometry and 3$\sigma$ upper limits
in Table~\ref{tab:hstphot}.

\begin{figure*}[!ht]
\includegraphics[width=0.92\textwidth]{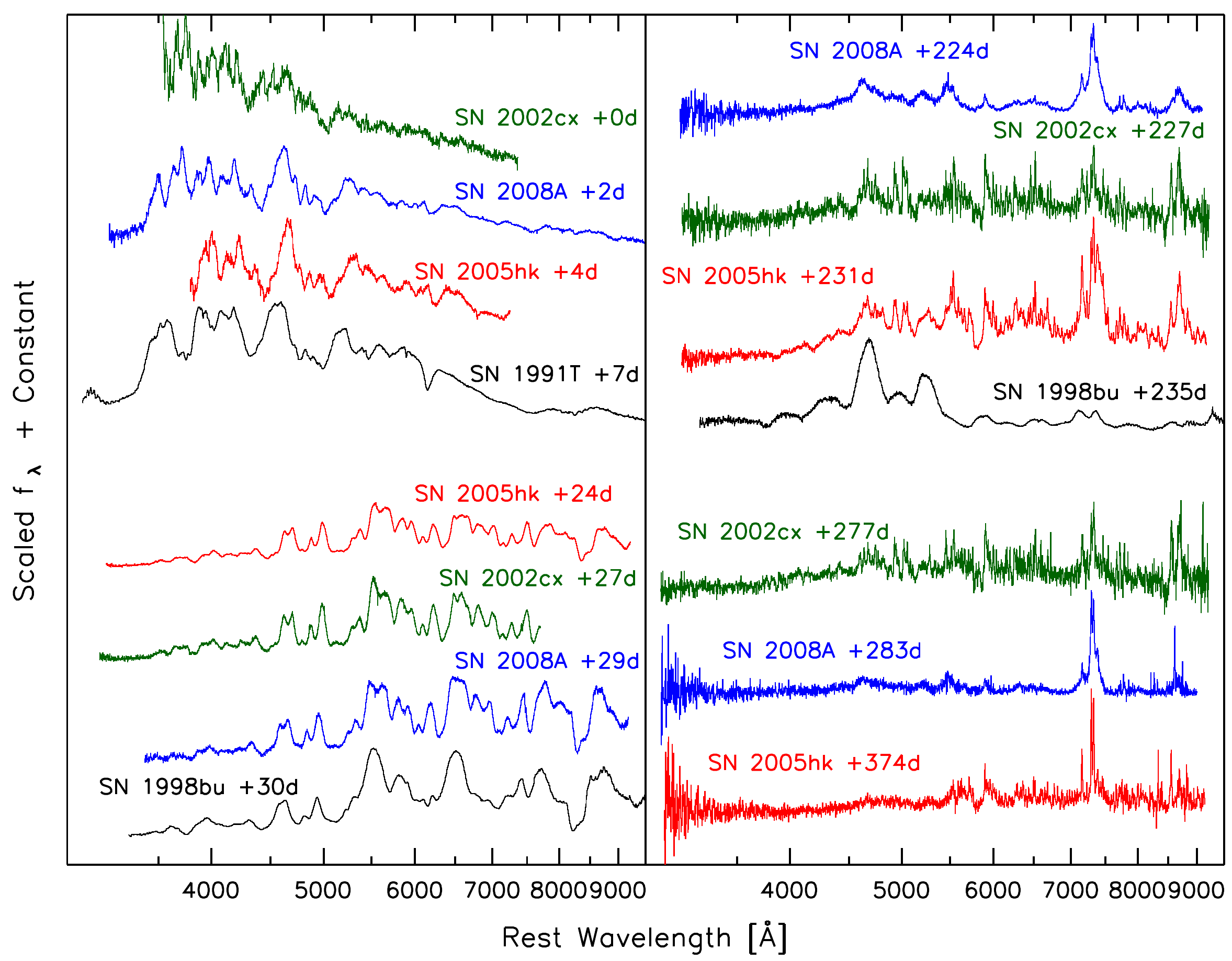}
\centering
\caption{Spectra of SN~2008A (this work) and SN~2005hk \citep[][and
    this work]{Phillips07} compared to those of SN~2002cx
  \citep{Li03,Jha06}. These three SNe~Iax have remarkably homogeneous
  spectra throughout their evolution, and diverge dramatically from
  normal SNe~Ia at late times, as shown by comparison spectra of
  SN~1991T \citep{Filippenko92T} and SN~1998bu
  \citep{Jha99,Li01pecrate,Silverman12}.}
\label{fig:spectra_compare}
\end{figure*}

\begin{figure*}[!ht]
\begin{center}
\includegraphics[width=0.92\textwidth]{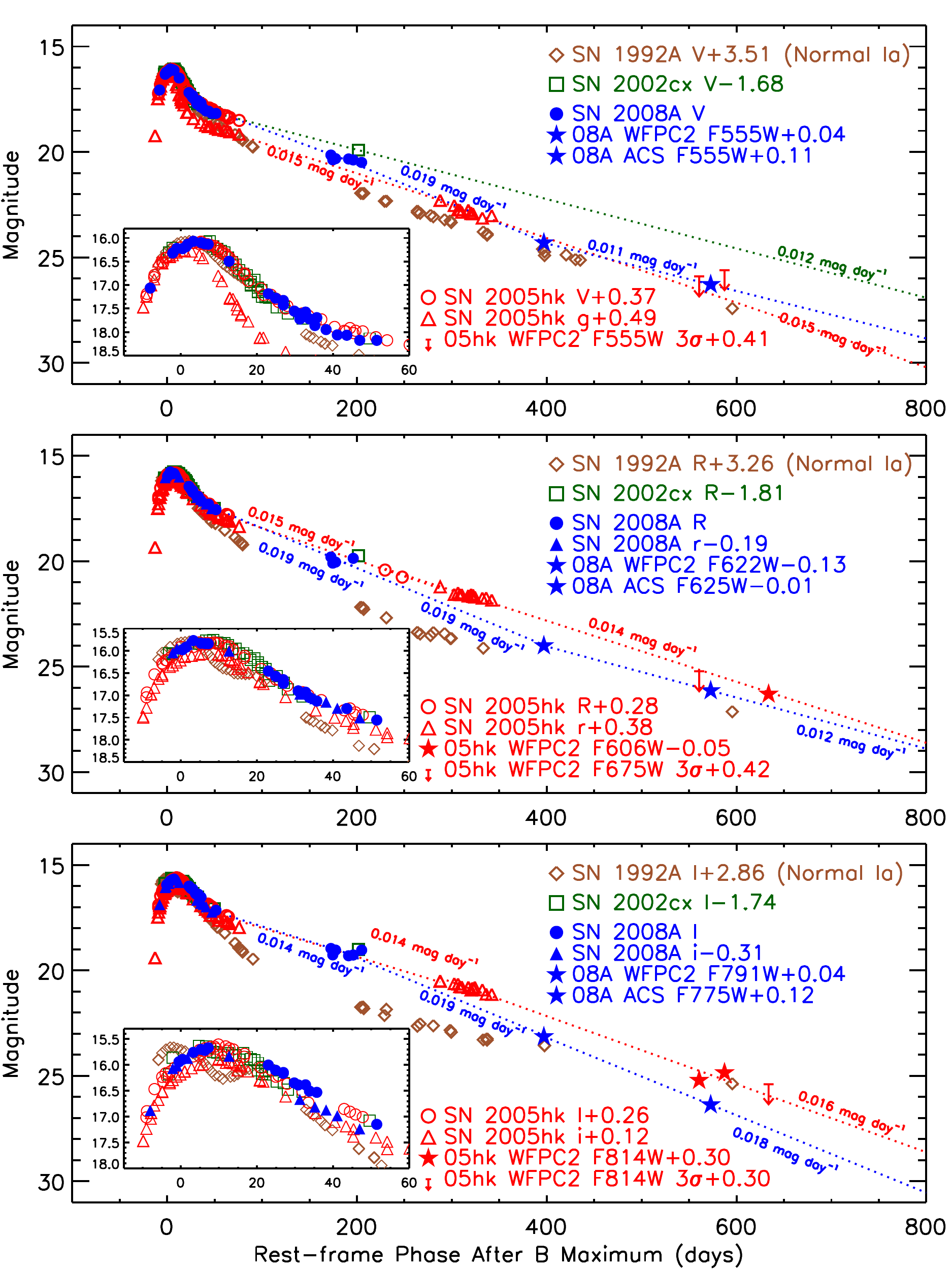}
\end{center}
\caption{Light curves of SN~2002cx \citep{Li03, Jha06}, SN~2005hk
  \citep[][and this work]{Phillips07,Sahu08}, and SN~2008A
  \citep[][and this work]{Ganeshalingam10, Hicken12}, compared to those of 
  the normal Type Ia SN~1992A \citep{Kirshner93}. These include both
  ground-based and {\it HST} observations as noted. The light curves
  have been shifted to match the peak of SN~2008A in $V$, $R$, and
  $I$. In each of the optical bands the decline rate is faster than
  the predicted $0.0098$ mag day$^{-1}$ for the decay of $^{56}$Co.}
\label{fig:figlate}
\end{figure*}

Both SNe~2005hk and 2008A are typical members of the SN~Iax
class. These two SNe along with SN~2002cx are quite homogeneous in
both their light curves and their
spectra. Figure~\ref{fig:spectra_compare} shows the spectroscopic
similarity of SNe~2002cx, 2005hk, and 2008A.  In
Figure~\ref{fig:figlate}, we plot photometry for SNe~2002cx
\citep{Li03, Jha06}, 2005hk \citep[][and this
  work]{Phillips07,Sahu08}, and 2008A, compared to the normal
SN~Ia~1992A \citep{Kirshner93}, all extending to late times. 

\section{Results}

\subsection{Lack of Near-infrared Secondary Maxima in SNe~Iax}

At early times, SN~2008A mimics the photometric behavior shown by
SN~2002cx and SN~2005hk, with broader light curves than SN~1992A,
particularly in the redder bands, as shown in the insets of
Figure~\ref{fig:figlate}. These SNe~Iax do not show a second maximum in
the near-infrared \citep{Li03,Phillips07}. This is puzzling if the
secondary maximum is ascribed to a transition in the dominant
ionization state \citep[\ion{Fe}{3} to \ion{Fe}{2};][]{Kasen06},
because SNe~Iax have hot early-time spectra with prominent
\ion{Fe}{3}, similar to SN 1991T-like SNe Ia that do show a strong
secondary maximum. Low luminosity SNe~Ia like SN~1991bg lack secondary
maxima \citep{Filippenko92bg,Leibundgut93}; in the ionization model
this is explained by an earlier transition onset in the cooler
spectra, merging the primary and secondary maxima. SNe~Iax could
undergo much more rapid cooling near maximum light than normal SNe~Ia
(to explain the \ion{Fe}{3} in the spectrum and yet lack of a
secondary maximum). This is plausible as SNe~Iax remain at high densities at late times instead of entering a nebular phase, which should enhance the cooling. However, we do not see evidence for rapid cooling in our late time observations, so perhaps an alternate model is needed.

\subsection{Spectral Features and Velocity Structure \label{sec:specfeatvel}}

One of the defining characteristics of SNe~Iax is the low velocity of
their spectral features. At early times, typical SNe~Iax have
expansion velocities for features such as \ion{Si}{2} of $\lesssim$
5000$\kms$, roughly half those seen in normal SNe~Ia \citep{Foley13},
though the overall appearance of the spectra is otherwise similar to
those of SN~1991T-like and SN~1999aa-like SNe~Ia. SNe~Iax at late
times are dominated by permitted lines from iron-group elements rather
than entering a nebular phase like normal SNe~Ia \citep{Jha06,
  Foley13}. Nonetheless, a few forbidden lines are
present, and the strongest features in late-time SN~Iax spectra are
the [\ion{Ca}{2}] $\lambda\lambda7291,7323$ doublet and [\ion{Fe}{2}]
$\lambda7155$ \citep{Jha06,Sahu08,Foley13}. Using these features, we
derive the redshift of the ejecta of SNe~2002cx, 2005hk, and 2008A to be $z = 0.02323$,
0.01176, and 0.01825, respectively. Ideally, we would use the redshift of the progenitor system, but this is not possible, so we adopt these redshifts to compare the shape of spectral features between objects. These emission lines have
unprecedentedly low velocities: in the last observed epoch, SN~2005hk
had velocities with a full-width at half maximum (FWHM) as small as $\lesssim 500 \kms$, the lowest ever measured for any
white-dwarf SN \citep[SN Ia or SN Iax;][]{Phillips07}. Some faint SNe IIP also have low velocities, within a factor of two of SN 2005hk at similar phase, but these typically do not get to a FWHM of $500 \kms$ until $\sim600$ days past maximum \citep{Maguire12}.

\begin{figure*}
\begin{center}
\includegraphics[width=1.0\textwidth]{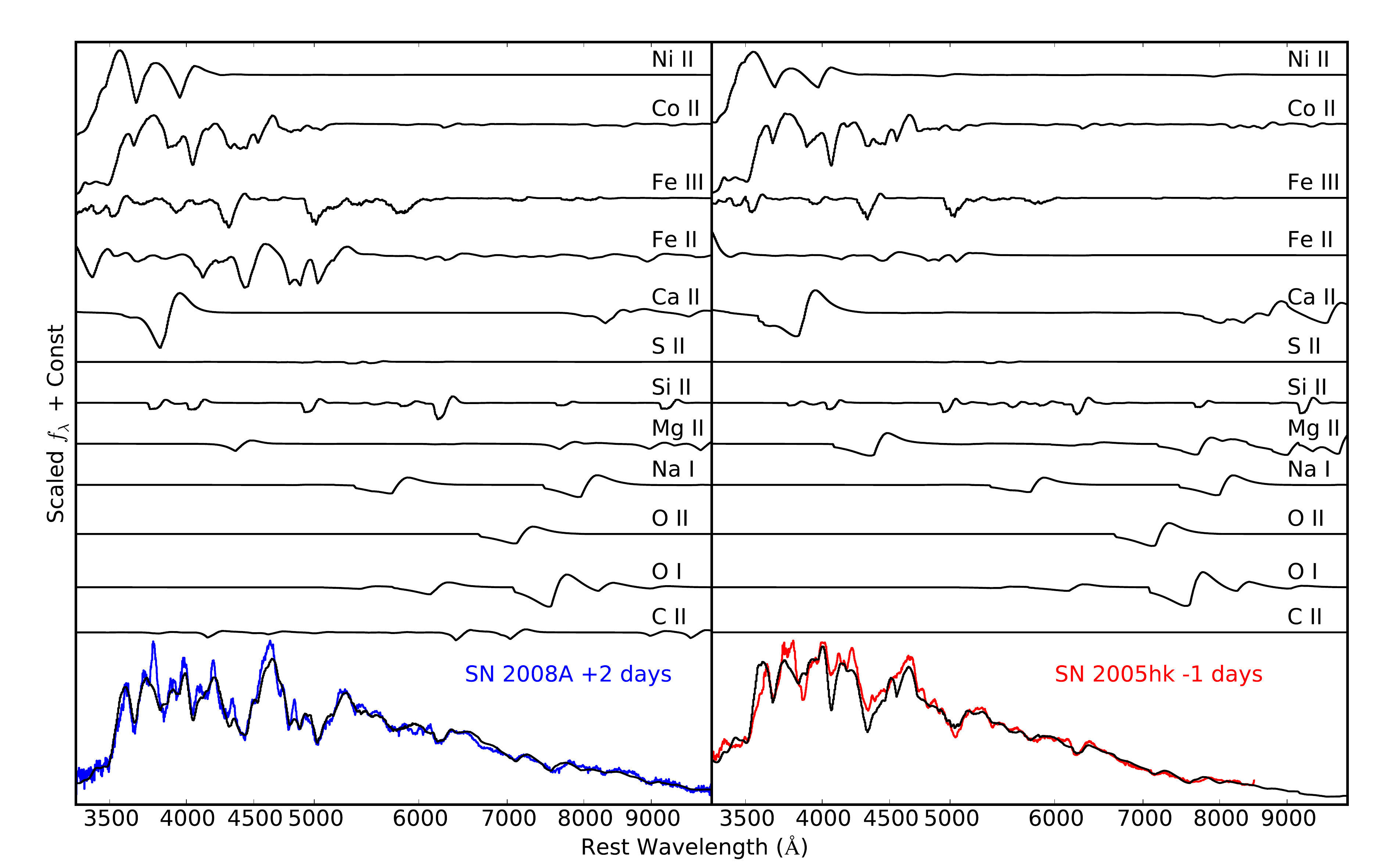}
\end{center}
\caption{Spectral fits for SN~2008A and SN~2005hk near maximum light. The observed spectra for SN~2008A for SN~2005hk are in blue on the left red on the right respectively. The best fit synapps model \citep{synapps} is overplotted in black and is decomposed by ion above. We find that the maximum light spectra are dominated by iron group elements, including \ion{Fe}{3}, but also includes features due to intermediate mass elements and (presumably) unburned carbon and oxygen. The best fit velocity for SN~2005hk is $\sim$7000$\kms$, about 1500$\kms$ lower than SN~2008A. This velocity difference persists at all epochs.}
\label{fig:synapps}
\end{figure*}

Figures \ref{fig:spec_05hk} and \ref{fig:spec_08a} show the full spectroscopic evolution of SN~2005hk and SN~2008A respectively. The features of SN~2005hk and SN~2008A match well at all epochs as illustrated in Figure \ref{fig:spectra_compare}. We used Synapps \citep[an optimizer for Syn++][]{synapps}, based on the SYNOW code \citep{Branch05}, to identify the strongest features in the photospheric spectra of SN 2005hk and SN 2008A and is shown in Figure \ref{fig:synapps}. Both objects are dominated by iron group elements, including \ion{Fe}{3} emission similar to 91T-like spectra but with significant contributions from intermediate mass elements. We also detect (presumably) unburned material in both objects: both SN~2005hk and SN~2008A have emission from \ion{O}{1} and possibly \ion{O}{2}. SN~2008A has strong carbon features (even relative to normal SNe~Ia), but SN~2005hk shows no evidence for carbon emission. One of the key differences between these two objects is that the velocities of the spectral features of SN~2008A are higher by 1500$\kms$ at maximum than those of SN~2005hk. The spectral features of SN~2005hk remain at lower velocities that those of SN~2008A at all epochs, even out to a year past maximum.  
 
\begin{figure*}[!ht]
\includegraphics[width=0.96\textwidth]{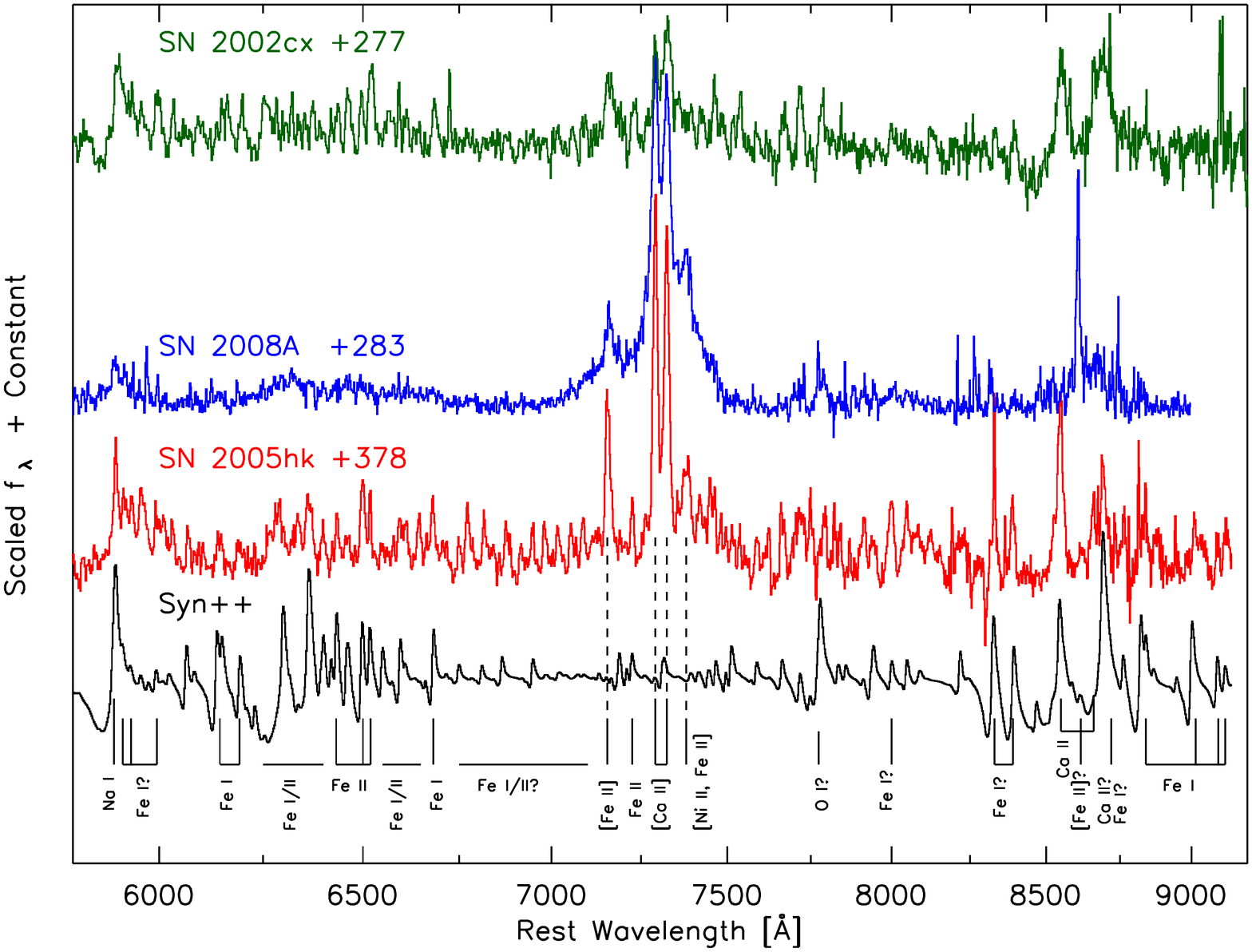}
\centering
\caption{Late-time spectra of SN~2002cx ($+$277 days; top, green),
  SN~2008A ($+$283 days; top middle, blue), and SN~2005hk ($+$378 days;
  bottom middle, red; \citealt{Sahu08}) compared to a synthetic model
  spectrum (black, bottom). The synthetic spectrum was created with
  Syn++ \citep[][similar to SYNOW \citep{Branch05}]{synapps}. The
  synthetic spectrum assumes Boltzmann excitation and only models the
  permitted lines. Non-LTE effects and forbidden lines are important for
  the relative strengths of the lines, but Syn++ is useful for the
  identification of the lines. We have marked the strongest forbidden emission features, discussed in the text, with dashed lines. The signal-to-noise ratio is
  substantially better in SN~2005hk, so this was used for the primary
  fit and then compared to SN~2002cx and SN~2008A. Line
  identifications are included under the synthetic spectrum.}
\label{fig:line_ids}
\end{figure*}

Figure \ref{fig:line_ids} compares the late-time spectra of SN~2002cx,
SN~2005hk, and SN~2008A.  While these spectra are qualitatively
similar, there are key differences (also analyzed in detail by
\citealt{Foley13}). At all epochs the velocities of SN~2008A are
higher than those of SN~2005hk, similarly to the photospheric spectra. SN~2002cx shows a stronger Ca near-IR
triplet than either SN~2005hk or SN~2008A. In the spectra of SN~2008A,
[\ion{Fe}{2}] $\lambda$7155 has an asymmetric profile.  This may be
caused by contamination from another line; there is a [\ion{Co}{1}]
feature at the wavelength in question, but no other [\ion{Co}{1}]
lines are seen (including lines more easily excited), making this
option unlikely.

\citet{Sahu08} identify [\ion{Fe}{2}] $\lambda 7389$ in their spectra
of SN~2005hk. However, there is another strong feature, [\ion{Ni}{2}]
$\lambda7378$, at almost the same wavelength. The feature near these
wavelengths is broader than the [\ion{Fe}{2}] $\lambda7155$ line, so
we argue that it is likely a blend of these two lines. There
is a strong line that is near [\ion{Fe}{2}] $\lambda8617$
in SN~2008A that is not seen clearly in SNe~2002cx or 2005hk, but is
observed in the normal SN~Ia~2003hv and has been used to measure
asymmetry in the inner layers of the ejecta \citep{Leloudas09}.

We used Syn++ \citep{synapps} to model the permitted lines in the late-time
spectra. Our results are also shown in Figure~\ref{fig:line_ids}. Because
the signal-to-noise ratio was the highest in the latest SN~2005hk
spectrum, we fit the lines in this spectra and then matched them to
the spectra of SNe~2002cx and 2008A. We find that \ion{Fe}{2} is
necessary to fit many of the lines between 6000 and 6400 \AA, as was
found by \citet{Jha06} and confirmed by \citet{Sahu08}.  While many
lines are fit well with \ion{Fe}{2}, some had remained
unidentified. We find that \ion{Fe}{1} significantly improves the fit
of the late-time SN~2002cx and SN~2005hk spectra.  There are strong
lines to the red of the P Cygni profile of \ion{Na}{1}
$\lambda5891$ that can be seen in all three objects (SNe~2002cx,
2005hk, and 2008A) that are well fit by \ion{Fe}{1} with an
excitation temperature near 5000~K. However, there are also
[\ion{Fe}{1}] lines that match these features, making the
identification of these lines ambiguous.

\begin{figure*}[!h]
\includegraphics[width=0.8\textwidth]{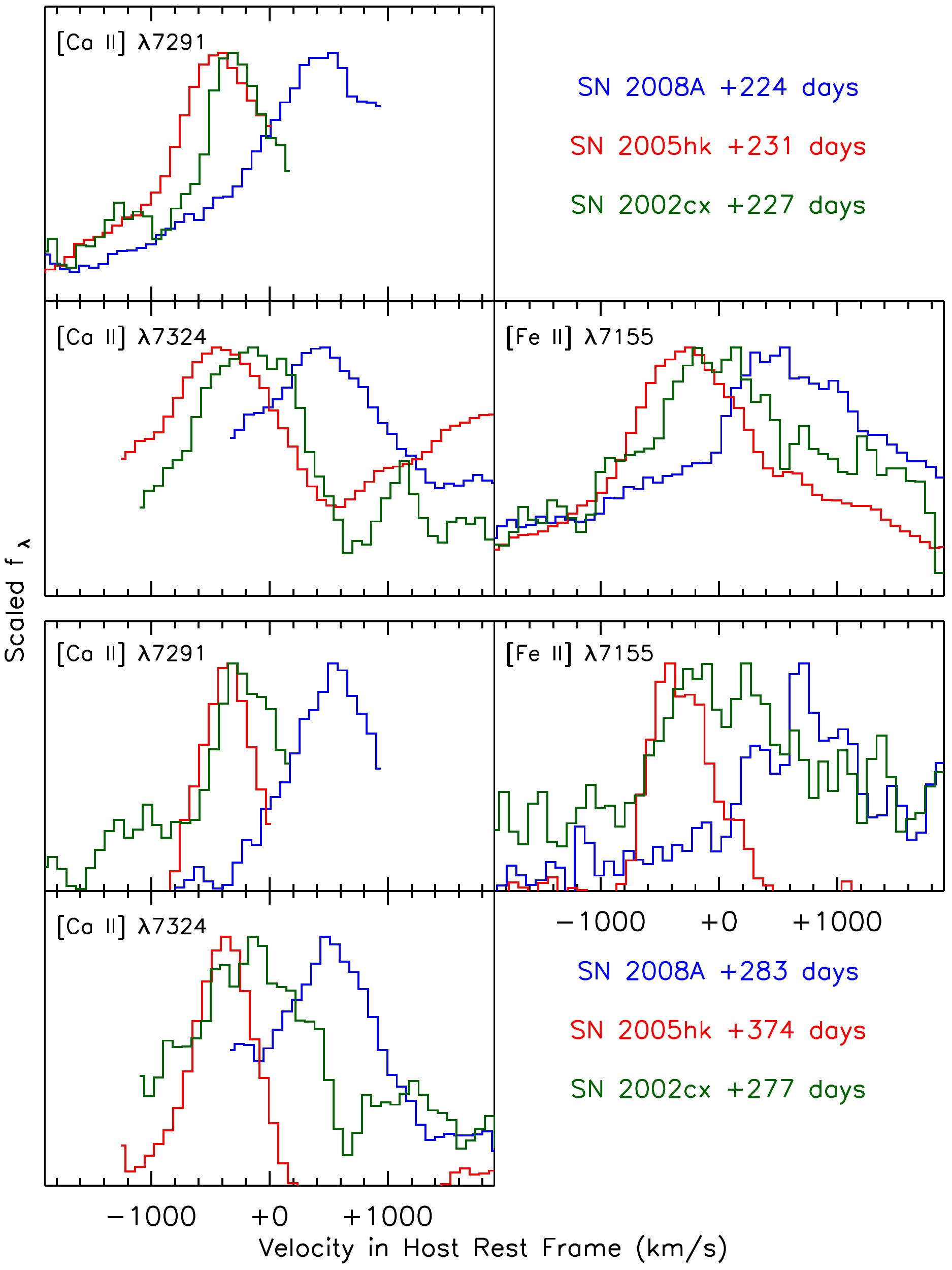}
\centering
\caption{Velocity structure of the forbidden lines [\ion{Ca}{2}]
  $\lambda7291$, [\ion{Ca}{2}] $\lambda7323$, and [\ion{Fe}{2}]
  $\lambda7155$ in SN~2002cx \citep[green;][]{Jha06}, SN~2005hk
  \citep[red;][and this work]{Sahu08}, and SN~2008A (blue; this
  work). The top three panels show each line at \about 230 days past
  $B$ maximum and the bottom three panels show the same lines at
  \about 300 days past $B$ maximum. The velocities of features are
  shown relative to the host-galaxy rest frame (in contrast to other
  figures, which are in the SN rest frame defined by these narrow
  lines). The host redshifts are $z = 0.023963 \pm 0.000087$ for
  SN~2002cx in CGCG 044$-$035 \citep{Meyer04,Wong06}, $z = 0.012993
  \pm 0.000041$ for SN~2005hk in UGC 272 \citep{Meyer04,Wong06}, and
  $z = 0.016428 \pm 0.000027$ for SN~2008A in NGC 634
  \citep{Theureau98}.  The line features of SN 2002cx and SN~2005hk
  are blueshifted compared to their hosts, while the features from
  SN~2008A are redshifted, with all the objects showing offsets of
  \about 400 $\kms$. The features also show a significant decrease in
  velocity width over time.}
\label{fig:velocity}
\end{figure*}

The velocity structure of the forbidden lines relative to the host galaxy is
interesting. \Cref{fig:velocity} displays the [\ion{Ca}{2}]
$\lambda\lambda7291,7323$ doublet and [\ion{Fe}{2}] $\lambda7155$ line
referenced to the host-galaxy (nucleus) rest frame (as opposed the rest frame of the ejecta used elsewhere in this work). As the figure
shows, the line velocities are largely consistent over time, though
the line widths decrease as the SNe evolve. \citet{Foley13} found that
the [\ion{Ca}{2}] and [\ion{Fe}{2}] features were shifted in opposite
directions relative to the host rest frame for the majority of
SNe~Iax, but for these three objects (also part of the
\citealt{Foley13} sample), we do not confirm this pattern, and find
consistent velocities from both [\ion{Ca}{2}] and [\ion{Fe}{2}].

In SN~2002cx, these lines were \emph{blueshifted} by $220 \pm 52
\kms$ relative to the SN host galaxy. In SN~2005hk, the lines were
also blueshifted by $370 \pm 22 \kms$. Contrarily, in SN~2008A the
lines were \emph{redshifted} by $547 \pm 10 \kms$. These velocity
shifts are significantly in excess of the host-galaxy rotation speeds,
which have maximum rotation velocities of $128 \pm 14 \kms$ (CGCG
044-035, host of SN~2002cx), $108 \pm 4 \kms$ (UGC~272, host of
SN~2005hk), and $225 \pm 7 \kms$ (NGC 634, host of SN~2008A),
respectively \citep{Paturel03}. This implies that the velocity shifts
are intrinsic to the SN explosion. We discuss the interpretation of
these results in Section \ref{sec:discussion}.

\subsection{Temperature and Density \label{sec:temp-dens}}

Important constraints for models are provided by the physical properties
of the ejecta, specifically the temperature and density at late
epochs. As discussed above, the late-time spectra of SNe~Iax are
dominated by permitted iron transitions which would imply that the
electron density has remained higher than in normal SNe~Ia. More
quantitatively, the ratio of the [\ion{Ca}{2}] doublet
($\lambda\lambda$7291, 7323) to the permitted \ion{Ca}{2} near-IR triplet
can be used to constrain the electron density and temperature
\citep{Ferland89}. As there is a degeneracy between temperature and
density in this method, to infer the electron density we need an
independent method for constraining the allowed temperatures.

\begin{figure*}[!ht]
\includegraphics[width=0.85\textwidth]{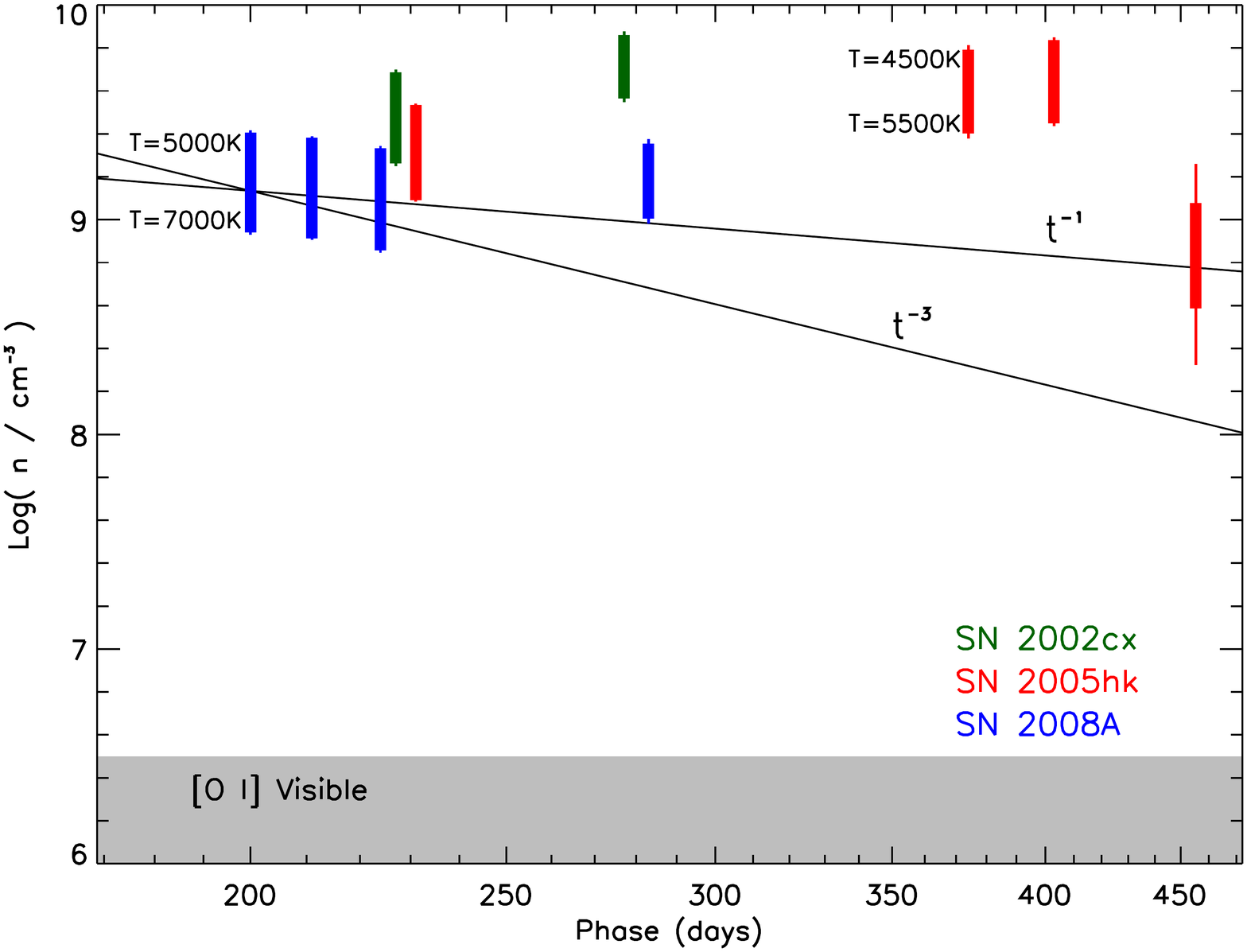}
\centering
\caption{Density evolution of SN~2002cx (green), SN~2005hk (red), and
  SN~2008A (blue) using the measured forbidden to permitted line ratio
  of \ion{Ca}{2}, based on the results of \citet{Ferland89}. The line
  ratio does not give a unique temperature and density, so the allowed
  temperature range was constrained by the presence of both
  \ion{Fe}{1} and \ion{Fe}{2} features, yielding the bands of allowed
  electron densities.  Each object has a electron density of $\sim
  10^{9}$~cm$^{-3}$ at \about 230 days past maximum light. The electron density
  is changing very slowly in all three objects; the results are
  consistent with no change in the electron density producing the
  observed emission. A density decrease proportional to $t^{-3}$ that would 
  be predicted by simple homologous expansion models is not consistent
  with the measured density evolution of these three SNe.
  }
\label{fig:density}
\end{figure*}

The presence of both \ion{Fe}{1} and \ion{Fe}{2} transitions in the
spectra of SN~2005hk is only allowed for a narrow range of
temperatures, if these lines arise from the same regions in the
ejecta, as suggested by the similar line profiles and velocities. The
Saha equation predicts a transition between \ion{Fe}{2} and
\ion{Fe}{1} transitions at \about 4500~K, and in local thermodynamic
equilibrium (LTE) calculations, \citet{Hatano99} find optical depth
exceeding unity in both \ion{Fe}{2} and \ion{Fe}{1} for $T \lesssim
7000$~K.  Although we do not expect equilibrium conditions, and
non-LTE effects are likely to be important, we can still use these
estimates as a reasonable range for the ejecta temperature, based on
the presence and strength of the \ion{Fe}{1} features among numerous
permitted lines with similar P Cygni line profiles and velocity
structure.  In Figure~\ref{fig:density}, we show the inferred
densities in SN~2002cx, SN~2005hk, and SN~2008A derived from the
\ion{Ca}{2} flux ratio for our allowed range of temperatures. We are
assuming that the iron and calcium are microscopically mixed; this is
plausible given their overlapping velocity ranges, but we cannot
definitively say they reside in the same physical region.

The spectra of all three of our objects are consistent with little
or no density evolution, and all have electron densities \about 10$^9$
cm$^{-3}$. Free expansion would predict that the density should go as
$t^{-3}$, but this is in clear contradiction with the
observations. Our measurements of the density are marginally
consistent with $t^{-1}$, but only if the temperature decreases
slowly. If the temperature decreases more rapidly (as might be
expected given the high densities, if radiative cooling dominates),
the density is likely evolving even more slowly than $t^{-1}$,
remaining roughly constant, or perhaps even increasing. One possible
explanation, given the decreasing line widths we observe (Figure
\ref{fig:velocity}), is that we are continuing to see the
``photosphere'' recede to ever lower velocities, and for some reason
the emitting region has a roughly constant density as the ejecta
dilute.

\subsection{Oxygen at Late Times}
\label{sec:oxygen}

\begin{figure*}[!ht]
\includegraphics[width=0.85\textwidth]{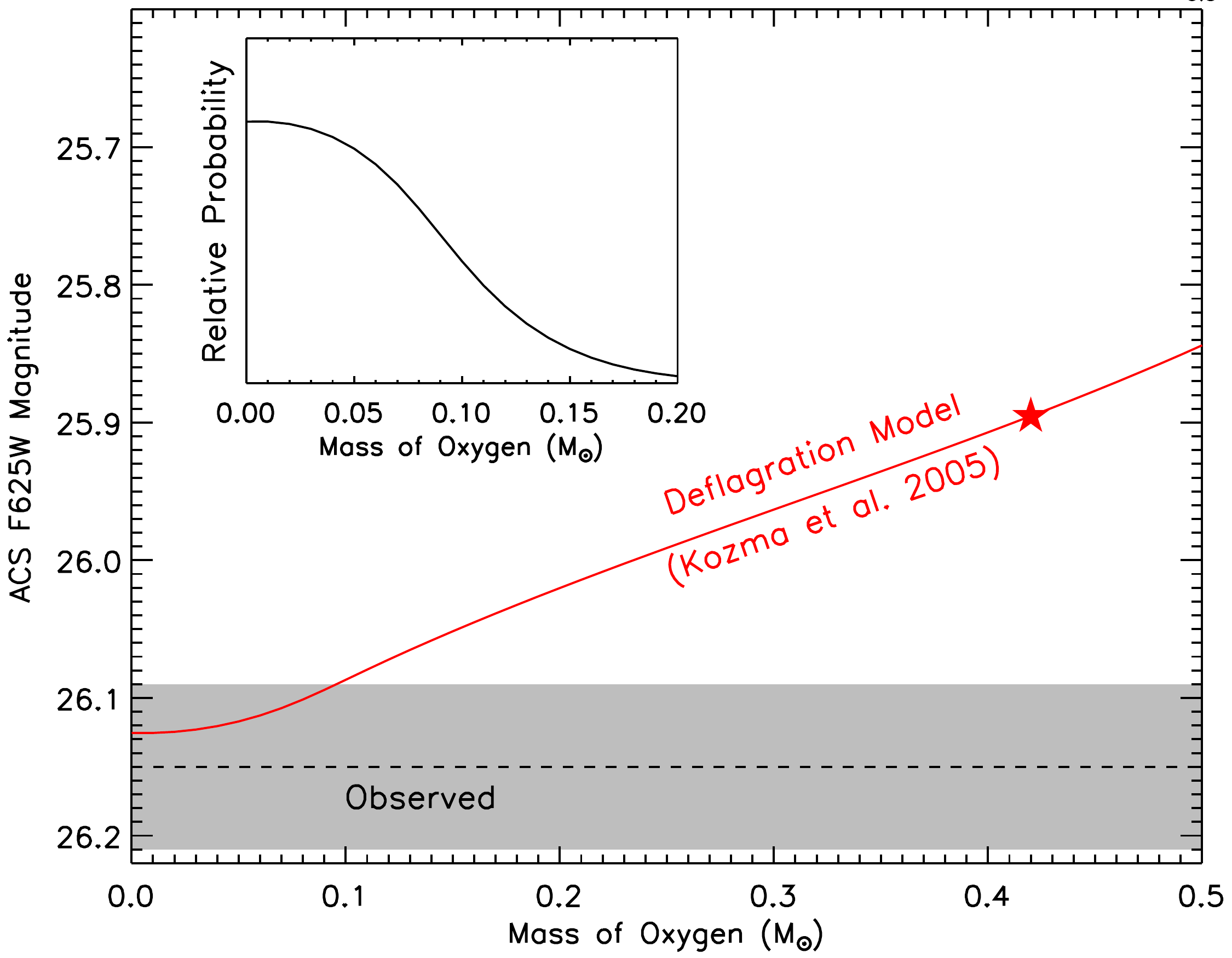}
\centering
\caption{Constraints on low-density oxygen from {\it HST}/ACS
  photometry of SN~2008A at $+$573 days past maximum light. The gray
  region represents the $1\sigma$ uncertainty in our measured
  photometry. The red line shows the predicted $r$-band magnitude
  (with flux from the [\ion{O}{1}] $\lambda6300$ line) assuming
  different amounts of unburned oxygen based on the models from
  \citet{Kozma05}. To calculate the probability density function (PDF)
  shown in the inset, we model the SED with a linear continuum and a
  Gaussian line profile centered at 6300 \AA. We calculate the
  likelihood of the measured optical photometry (in F555W, F625W, and
  F775W) and marginalize over the slope and normalization of the
  continuum to derive our final PDF. The PDF peaks at zero line flux,
  and puts a $95\%$ confidence upper limit of 0.14 $M_\odot$ of oxygen
  below the critical density for [\ion{O}{1}] $\lambda6300$. The pure
  deflagration model of \citet{Kozma05} predicts 0.42 $M_\odot$ of
  low-density unburned oxygen, which is ruled out at $> 5\sigma$.}
\label{fig:late_o}
\end{figure*}

As described in Section \ref{sec:introduction}, pure deflagration
models have been raised as a possibility to explain SN~Iax
explosions. These models generically predict a turbulent burning front
which causes strong mixing in all layers of the ejecta, implying there
should be unburned material (presumably C and O) in the innermost
layers of the ejecta \citep{Gamezo03}. At late epochs, one of the
strongest spectral features should be [\ion{O}{1}] $\lambda6300$ (assuming the entire white dwarf is disrupted). 
\citep{Kozma05}. 

There is no evidence of this line in the late-time nebular spectra of
normal SNe~Ia\footnote{\citet{Taubenberger13} have recently shown the
  evidence for nebular [\ion{O}{1}] $\lambda\lambda$6300,6363 with a
  complex line provide in a late-time spectrum of SN~2010lp, a
  subluminous SN 1991bg-like supernova.}. Instead, the nebular spectra of
normal SNe~Ia are dominated by forbidden transitions of iron-peak
elements. The favored explanation for this is that the burning front
transitioned from a subsonic deflagration to a supersonic detonation
\citep{Khokhlov91, Gamezo05}, which subsequently burns the central
material uniformly to the iron peak.

SNe~Iax can have strong carbon features before and at maximum light (like SN~2008A as discussed above),
corresponding to unburned material in the outer layers of the ejecta
\citep[][and references therein]{Foley13}. Some normal SNe~Ia also
show these lines at early times, but if present, they usually
disappear by maximum light \citep{ Thomas07, Thomas11, Parrent11,
  Folatelli12, Silverman12_carbon}. This may imply that unburned
material is present at higher mass fractions deeper into the ejecta in
SNe~Iax compared to normal SNe~Ia.  If SNe~Iax are to be explained as
pure deflagrations (no transition to supersonic burning), we expect to
see evidence of unburned material at all velocities, including the
central regions that are revealed at late times.

\citet{Jha06} tentatively identified \emph{permitted} \ion{O}{1}
$\lambda7774$ in SN~2002cx at 227 and 277 days after $B$ maximum. As
can be seen in Figure~\ref{fig:line_ids}, the same feature is present
in SN~2008A, but even stronger. If the identification is correct,
this matches the model predictions of a pure deflagration explosion
that never transitioned to a detonation. However, this feature is a
\emph{permitted} transition, implying that the density of the
ejecta is unexpectedly high out to \about 280~days past maximum. By
this phase, SNe~Ia have transitioned to a nebular phase dominated by
forbidden transitions of iron-peak elements\footnote{\citet{Branch08}
  argue that permitted lines dominate optical spectra of normal SNe~Ia
  as late as a few months past maximum light, though by about 160 days
  past maximum typical SNe Ia have nebular spectra
  \citep{Silverman13}.}. If the identification of \ion{O}{1}
$\lambda7774$ is correct and there is unburned oxygen at low velocity,
[\ion{O}{1}] $\lambda 6300$ should be a prominent feature in the
nebular spectra of these objects. We searched for evidence of
[\ion{O}{1}] $\lambda 6300$ in spectra of SN~2005hk taken \about
400~days after maximum, but SN~2005hk did not enter a nebular phase
even at these late epochs and there was no evidence for [\ion{O}{1}]
$\lambda 6300$.

Spectroscopy is no longer feasible after these late epochs because the
SN is too faint.  Instead, we use photometry to constrain the strength
of the [\ion{O}{1}] line at $+$573 days after maximum. [\ion{O}{1}]
$\lambda6300$ is near the center of the $r$ band (F625W) and is
reasonably isolated from any other spectral features expected to be
present.  Therefore, we use the $r$-band photometric flux as a proxy
for the flux in the oxygen line. If the [\ion{O}{1}] $\lambda 6300$
line began to dominate other spectral features and the nearby
continuum, as is predicted by the pure deflagration models
\citep{Kozma05}, we would expect a strong $r-i$ and $V-r$ color change
as the SN transitions to a nebular phase.

To measure this color change, we examine two epochs of {\it HST}
photometry of SN~2008A: the first epoch is at +396~days, at which we
expect no contribution from [\ion{O}{1}] $\lambda6300$ based on our
spectra of SN~2005hk at similar epochs. In our second observation at
+573~days, we do see a strong $r-i$ color change, but $V-r$ remains
relatively unchanged (see the discussion in Section \ref{sec:ir-catastrophe}); 
this cannot be easily explained by just the
appearance of a strong line in the $r$ band.

Our photometry of SN~2008A at $+$573 allows us to quantitatively
constrain the amount of oxygen in the ejecta below the critical
density of $10^{6.5}$ cm$^{-3}$. The complete deflagration models of \citet{Kozma05} give us
a relationship between the oxygen mass and the [\ion{O}{1}]
$\lambda6300$ line flux. If we unrealistically assume that \emph{all}
of the observed flux in F625W is from an oxygen line, we would
derive a mass of 0.63 $M_\odot$ of oxygen at low density. In reality,
the oxygen line flux is only part of the observed broad-band F625W
photometry. If we extrapolate just the $r$-band photometry from
earlier times to day 573, the data allow for only 0.40 $M_\odot$ of
oxygen to contribute additional line flux.

However, we can derive much more stringent oxygen mass limits if we
constrain the SN spectral energy distribution (SED) in this wavelength
region. Based on the observed spectroscopy at earlier times, we see
that SNe~Iax remain pseudo-continuum dominated in broad-band
photometry to \about 400 days past maximum, and the subsequent
photometry does not show dramatic changes in late-time behavior among
the different passbands. This implies that we can use our measured
F555W and F775W flux to bracket and constrain the expected continuum
flux in F625W, with any new oxygen line flux showing up as an excess.

To do this we model the SN SED as a linear continuum (in logarithmic
wavelength; $f_\lambda \propto \log \lambda$) plus a Gaussian line
profile at 6300~\AA\ with a width of 500$\kms$, with a line flux
calibrated to the oxygen mass as above \citep{Kozma05}. We perform
synthetic photometry on the resulting spectra to compare to our {\it
  HST}/ACS observations in F555W, F625W, and F775W. Essentially, the
F555W and F775W data determine the continuum normalization and slope,
and the F625W photometry then constrains the oxygen line flux. Our
broad-band measurements are insensitive to the exact shape of the line
profile, and using the three bands together, we marginalize over the
nuisance parameters of the continuum slope and normalization to derive
the oxygen mass limit.

Our constraint is shown in Figure~\ref{fig:late_o}; we find that the
mass of oxygen below the critical density of [\ion{O}{1}]
$\lambda6300$ is $< 0.14$~$M_{\odot}$ at $95\%$ confidence.  Complete
deflagration models that fully disrupt the white dwarf predict \about 0.4~$M_{\odot}$ of unburned
oxygen \citep{Kozma05}. This is ruled out at high confidence ($> 5\sigma$). Either a
large amount of unburned material is not present in SNe~Iax, or it
remains at high density to very late epochs. Even if the ejecta mass
were as low as \about 0.5~$M_{\odot}$, we would expect \about
0.15~$M_{\odot}$ of unburned oxygen (assuming the same ratio of oxygen mass to ejecta mass as the models from \citealt{Kozma05})
--- and this would still be ruled out at $> 95$\% confidence.

As discussed, our results depend on the assumed shape of the SN SED.
However, if we enforce a smooth continuum over the observed wavelength
range (so we can incorporate the F555W and F775W data into our
analysis), different prescriptions for the exact continuum shape lead
to only small changes in the oxygen mass constraint of \about 0.02
$M_\odot$.

Based on our temperature and density measurements, the electron density may have been
significantly higher than the critical density for [\ion{O}{1}]
$\lambda6300$ of $10^{6.5}$ cm$^{-3}$ during our latest {\it HST}
measurement of SN~2008A, nearly 600~days past maximum brightness. But
if the roughly constant observed density is somehow the result of a
receding photosphere, this would imply a significant amount of
material at higher velocities and lower densities, and any oxygen
mixed there should be detectable in [\ion{O}{1}] emission as probed by
our photometry. Complete deflagration models such as those of \citet{Kozma05} predict
strong mixing, so we disfavor the idea that all the oxygen is at
low velocity, shielded by a photospheric ``iron curtain'' at higher
velocity \citep{Leonard07}. \citet{Sahu08} find in their models of SN~2005hk that if a significant fraction of the gamma rays are still exciting photospheric transitions, the resulting [\ion{O}{1}] $\lambda6300$ is weak enough at +250 days to be consistent with the nondetection in the observations, even in the case of 0.8 $M_\odot$ of unburned oxygen. However, it is unclear whether this can be sustained as late as +600 days, where our observations of SN 2008A otherwise significantly constrain any low-density oxygen. It is possible that this points to a diversity between objects like SN~2005hk and SN~2008A, but even for SN 2005hk itself, the suppression of [\ion{O}{1}] emission must continue until at least +456 days past maximum, as it is undetected in our latest spectrum. Our conclusion is that either the
spectroscopic detection of permitted \ion{O}{1} $\lambda$7774 at low
velocity in SNe~2002cx and 2008A (as seen in Figure~\ref{fig:line_ids}) is a misidentification, or, if not, a more complicated model (perhaps like one that produces a bound remnant) must be
developed to explain the apparent contradiction between the permitted
and forbidden lines; we discuss further implications of this finding in 
Section \ref{sec:discussion}.

\subsection{IR Catastrophe \label{sec:ir-catastrophe}}

\begin{figure*}[!ht]
\includegraphics[width=0.6\textwidth]{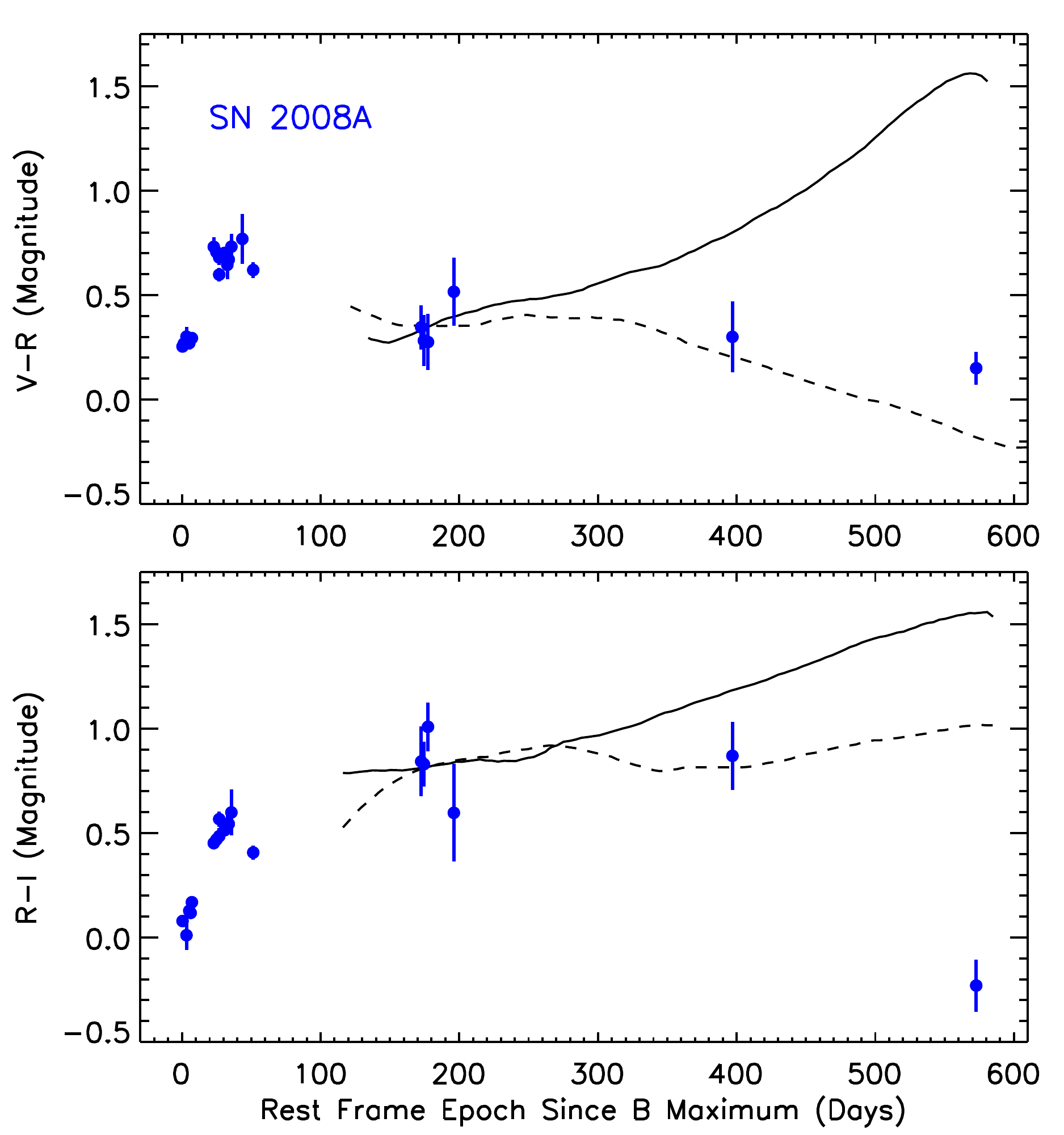}
\centering
\caption{$V-R$ (top) and $R-I$ (bottom) color evolution of SN~2008A
  compared to IR-catastrophe models from \citet{Sollerman04}. The
  dashed line includes photodisintegration in the model while the
  solid line does not. The data have been standardized to $V$, $R$,
  and $I$ as in Figure \ref{fig:figlate}. The models have been shifted
  vertically to match the data at \about 200 days.}
\label{fig:sollerman_colors}
\end{figure*}

The ``infrared catastrophe'' is a long-predicted, generically expected
phenomenon that has nonetheless never been observed in normal SNe~Ia
\citep{Leloudas09}\footnote{\citet{Taubenberger11} suggest that the IR
  catastrophe could be at play in the rapid late-time decline of the
  light curve of the ``super-Chandrasekhar''
  SN~2009dc.}. \citet{Axelrod80} calculated that once the ejecta
temperature dropped below a critical value \about 1500~K, there would
be a radical redistribution of energy from the optical to the far-IR,
dominated by fine-structure lines of iron like [\ion{Fe}{1}]
24~$\mu$m and [\ion{Fe}{2}] 26~$\mu$m rather than the forbidden
iron-peak lines seen in the optical and near-IR that dominate the
SN~Ia bolometric luminosity a few hundred days past maximum light
\citep{Sollerman04}. This redistribution of energy has been used to
explain the line emission of SN~1987A \citep{Kozma98}, but it has never
been observed in a normal SN~Ia. One possible explanation is that the ejecta
of SNe~Ia stay above this critical temperature until after they are
too faint to observe. \citet{Leloudas09} find that even 785~days after
$B$-band maximum brightness, SN~2003hv is not compatible with an
IR catastrophe.

SNe~Iax are unique testing grounds for this phenomenon. Because these
objects remain at high density for so long, we might expect them to
have enhanced cooling compared to normal SNe~Ia. This could cause the
IR catastrophe to happen early enough that it would still be
feasible to image the SN. Qualitatively, if we use the excitation
temperature of the iron (and assume microscopic mixing), we can
estimate what the temperature of the ejecta will be at the time of our 
final {\it HST} measurement. Constructing a simple model that assumes that 
the excitation temperature of iron is \about 10,000~K at maximum brightness
(giving rise to the dominant \ion{Fe}{3} lines observed) and \about
4000~K at 300~days after maximum brightness (see Section 
\ref{sec:temp-dens}), it is plausible (though by no means
required) that the temperature at the epoch of our final {\it HST}
measurement could approach the \about 1500~K threshold for the
IR catastrophe. \citet{Sollerman04} model this transition and
find that the most easily observable signature (i.e., not in the
far-IR) for the IR catastrophe is a dramatic color change
in the near-IR, but we only have one near-IR measurement, ($J$:F110W), so we
cannot observe this color change directly.

\citet{Sollerman04} also show that the optical colors should become
significantly redder in all optical bands during and after the IR
catastrophe, with significant $V-R$ and $R-I$ color changes of \about
1~mag. We compare this model with the observations of SN~2008A in
Figure~\ref{fig:sollerman_colors}. Though there is some freedom to
shift to model predictions vertically in magnitude and horizontally in
time on the plot, we nonetheless see that the colors of SN~2008A
change in the \emph{opposite} way of the predicted color
changes. Rather than getting redder, $V-I$ and $R-I$ data are bluer at
late times (with $V-R$ remaining basically unchanged). Thus, we find no
evidence for an IR catastrophe in SN~2008A, out to nearly 600
days past maximum light, despite the high densities.

The late-time spectra of SNe~Iax differ significantly from those of 
normal SNe~Ia. Normal SNe~Ia cool through forbidden iron lines in the
nebular phase. SNe~Iax have not been shown to ever become nebular, and
the strongest forbidden lines are from [\ion{Ca}{2}]. There are iron
features in the late-time spectra of SNe~Iax, but a significant
fraction of the emergent radiation is produced by permitted
transitions. It is not even clear that radiative cooling dominates
over adiabatic expansion at late times in SNe Iax. Because of these
differences, it is possible that the predictions of IR catastrophe
models for normal SNe~Ia do not apply to SNe~Iax. Ironically, it could
be that the high densities that should enhance the cooling and lead to
an earlier IR catastrophe are instead responsible for quenching the
expected changes in forbidden-line emission.

\subsection{Bolometric Luminosity}

We use ground-based and {\it HST} photometry to estimate the
UV/optical/near-IR luminosity $L_{\rm UVOIR}$ of SN~2002cx,
SN~2005hk, SN~2008A, and the normal SN~Ia~2003hv. \citet{Sollerman04}
and \citet{Leloudas09} show that the flux in the near-IR becomes a
significant fraction of the bolometric luminosity of SNe~Ia at late
times, and we are able to replicate the near-IR corrections of
\citet{Leloudas09} in our calculations for SN~2003hv.

For SNe~2002cx and 2008A, we do not have much information about the
flux in the IR. \citet{Phillips07} present near-IR light
curves of SN~2005hk to a few months past maximum light, showing
significantly enhanced near-IR flux compared to normal SNe~Ia. However, at
these epochs the near-IR contribution to the bolometric luminosity is only
a few percent, and it is unclear whether the excess near-IR flux persists
to later epochs, or merely becomes more significant earlier in SNe~Iax
than in normal SNe~Ia.

To correct the SNe~Iax for the poorly constrained near-IR flux, we
integrate our interpolated photometric SED from 3000~\AA\ to
10000~\AA\ and add the IR contribution fraction measured for SN~2003hv
\citep{Leloudas09}. This approach produces results that are consistent
with the late-time near-IR data we do have (a WFC3/IR F110W measurement
for SN~2008A and NICMOS F110W and F160W upper limits for SN~2005hk;
Table \ref{tab:hstphot}). For the final two {\it HST} observations of
SN~2005hk, we have a detection only in a single optical photometric
band; to calculate the bolometric luminosity at these late epochs, we
use the SED of SN~2008A at a similar phase and scale the flux to match
the SN~2005hk observation.

\begin{figure*}[!ht]
\includegraphics[width=0.85\textwidth]{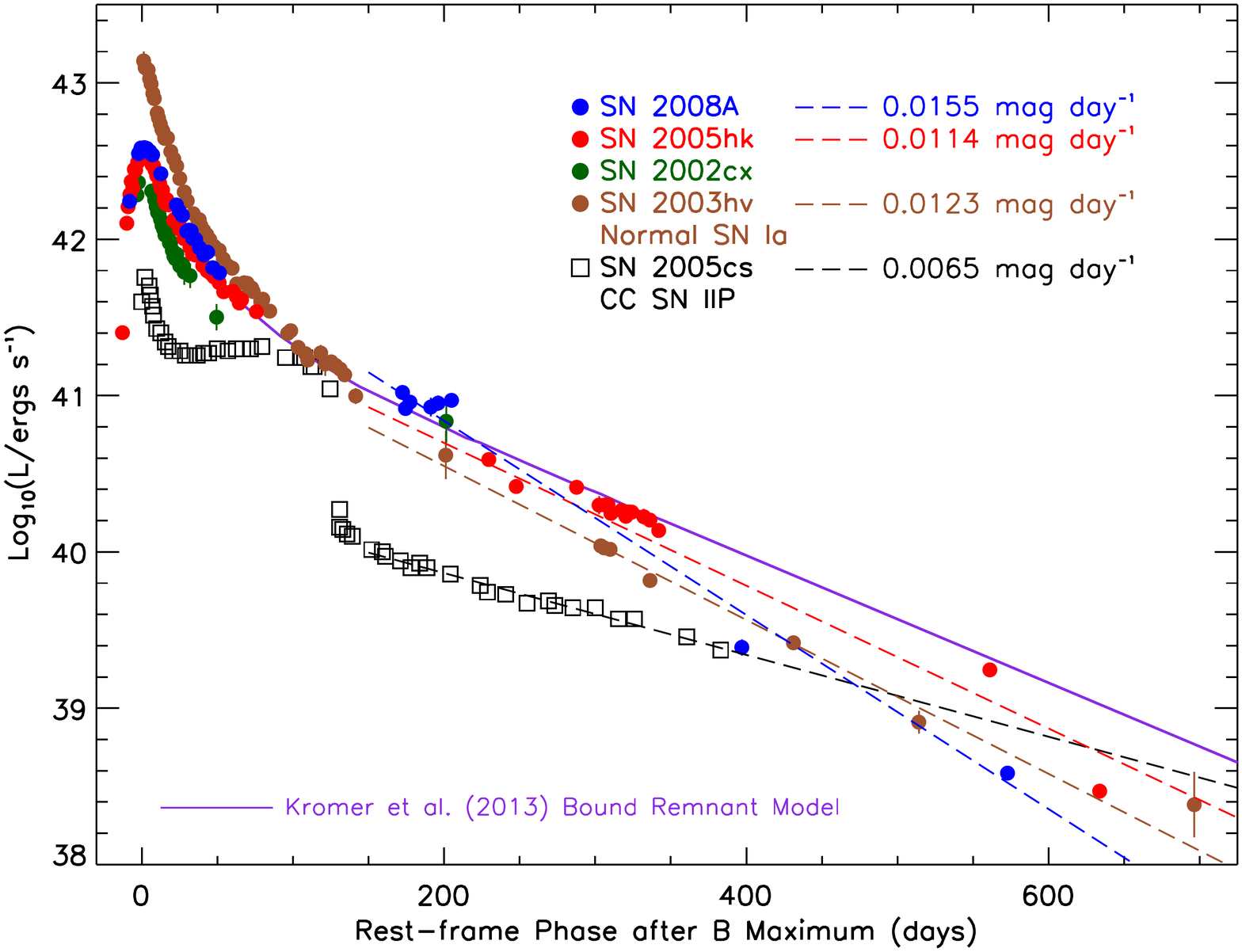}
\centering
\caption{Bolometric light curves of SN~2002cx \citep{Li03,Jha06},
  SN~2005hk \citep[][and data from SDSS-II]{Phillips07,Sahu08},
  SN~2008A \citep[][and this work]{Ganeshalingam10,Hicken12}, the
  normal Type Ia SN~2003hv \citep{Leloudas09}, and the Type IIP SN
  2005cs \citep{Pastorello09}. The bolometric luminosities of SNe~Iax
  are similar to those of normal SNe~Ia starting a few months after 
  maximum, but differ significantly from those of SNe~IIP. 
  SN~2003hv declines much
  faster than SN~2008A and its counterparts at \about 200 days past
  maximum. The magenta line shows the model from \citet{Kromer13} that
  assumes complete gamma-ray trapping. This model fits most of the
  data well, but it is not consistent with the latest measurements of
  SN~2008A and the last measurement of SN~2005hk. }
\label{fig:bolometric}
\end{figure*}

As shown in Figure~\ref{fig:bolometric}, we find that the bolometric
light curves of SNe~Iax are markedly different from that of a typical
core-collapse SN~IIP, SN~2005cs \citep{Pastorello09}, though of course
core-collapse SNe show a wide range of behavior.  The SNe~Iax are
similar to each other, and significantly less luminous than the
normal SN~Ia 2003hv at maximum light, but broadly similar to SN~2003hv
starting \about 100 days after maximum light.

The faster decline rate of SN~2003hv compared to SNe~Iax for the first
200~days suggests that the escape fraction of gamma-rays in normal
SNe~Ia increases more quickly than in SNe~Iax. This may result from a
more rapid decrease in the density of the ejecta in normal SNe~Ia
(also suggested by our inferred SNe~Iax densities; Figure
\ref{fig:density}). The bolometric luminosity of SN~2005hk roughly
follows the full gamma-ray trapping model of \citet{Kromer13} for
all but the latest epoch. SN~2008A is also consistent with the full
gamma-ray trapping model at early times.

However, in the latest observations, more than \about 400 days past
maximum, that trend reverses: SN~2003hv declines more slowly than
SN~2005hk or SN~2008A. At these epochs, the decline rate for SN~2003hv
is close to the predicted $^{56}$Co to $^{56}$Fe decay rate of
0.0098~mag day$^{-1}$. This may be an indication of \emph{positron
  trapping} in the ejecta of this normal SN~Ia. About 3\% of the
$^{56}$Co decay energy goes to positrons, compared to 97\% in
gamma-rays \citep{Milne99}. Even if the gamma-ray escape fraction is
near unity, positrons may be trapped by (nonradial) magnetic fields,
and can lead to a shallowing of the bolometric luminosity decline rate
when the luminosity falls to \about 3\% of the full gamma-ray trapping
prediction.

At the epochs of our last observations of both SNe~2005hk and 2008A,
the bolometric luminosity drops below that of the \citet{Kromer13} model. 
This could be a sign that the gamma-ray escape fraction is beginning to
increase, at a much later epoch than occurs for normal SNe~Ia because
of the much higher densities in the ejecta. We might then predict that
at even later times, again when the bolometric luminosity falls to
\about 3\% of the full gamma-ray trapping prediction, the bolometric
light curves of SNe~Iax may flatten because of positron
trapping. Nonetheless, this remains speculative, as we are not
measuring the true bolometric luminosity. If either normal SNe~Ia or
SNe~Iax undergo an IR catastrophe, the UVOIR luminosity will not trace
the bolometric luminosity, complicating efforts to constrain the
positron trapping. Moreover, many normal SNe~Ia show steeper decline
rates at late epochs, inconsistent with full positron trapping
\citep[e.g.,][]{Lair06}, and in some cases even when the near-IR flux is 
included in the ``bolometric'' luminosity \citep{Stanishev07_03du}.

\section{Discussion}
\label{sec:discussion}

As described in Section \ref{sec:specfeatvel}, the velocity shifts measured
from late-time forbidden lines in SNe~2002cx, 2005hk, and 2008A are
larger than can be accounted for by galactic rotation alone.  These
offsets are measured from [\ion{Ca}{2}] and [\ion{Fe}{2}] lines; if
the ejecta were optically thin in these lines as expected, the
velocity offsets could be caused by bulk velocity shifts of the
ejecta. This might correspond to the high-velocity remnant predicted
by the failed deflagration model of \citet{Jordan12}, although
significant velocity kicks are not found in similar models from
\citet{Kromer13} or \citet{Fink14}.

The velocity shifts could also be the result of an asymmetric
explosion. However, \citet{Chornock06} measured less than 1\%
polarization for SN~2005hk, implying very little asymmetry at early
times \citep[see also][]{Maund10}. For the normal Type Ia SN~2003hv,
\citet{Leloudas09} measured a velocity shift of \about 2600$\kms$ in
the [\ion{Fe}{2}] $\lambda8617$ line. \citet{Maeda10} show that the
large velocity shifts in the inner regions of SN~2003hv could arise
in the deflagration phase of a delayed detonation model.
\citet{Maeda10_2} use a model with asymmetry of the deflagration phase
to explain the variation of SN~Ia velocity gradients, and show that
different viewing angles can account for much of the diversity of
SNe~Ia \citep[but see][who find a host-galaxy dependence that cannot
  be explained by viewing angle alone]{Wang13_science}.  In this
model, the deflagration stage is characterized by turbulent burning
underlying a convective, bipolar structure which creates an asymmetry
in the ejecta \citep{Kuhlen06,Ropke07}. The velocity shifts for
SNe~2005hk and SN~2008A are much smaller than that for SN~2003hv,
which might be caused by a lower kinetic energy during the
deflagration phase, and perhaps implying less total burning. The fact that the forbidden-line velocity \textit{offsets} are comparable to line \textit{widths} in both SNe Iax and normal SNe Ia may point to a common physical origin, though this may be generic to all SNe. 

The identification of \ion{S}{2} and \ion{Si}{2} lines in early-time
spectra of SNe~Iax point to a thermonuclear event \citep{Foley10,
  Foley13}. Our identification of late-time [\ion{Ni}{2}]
$\lambda7378$ in SN~2005hk and SN~2008A could thus have important
consequences, as \citet{Maeda10,Maeda10_2} identify both [\ion{Fe}{2}]
$\lambda7155$ and [\ion{Ni}{2}] $\lambda7378$ as lines that correspond
to {\it deflagration ashes}. They argue that these lines trace
material subject to long exposure to low heat, as opposed to
[\ion{Fe}{3}] $\lambda4701$ that traces the detonation phase. This
[\ion{Fe}{3}] transition is not obvious in SN~Iax spectra at late
times, which may imply that there is no transition to a detonation.

What could suppress a transition to a detonation in SNe~Iax? From
simulations, \citet{Ropke07} suggest that the number and position of
deflagration ignition points could be a significant factor.  For
example, they claim that a model with two overlapping ignition bubbles
at the center of the white dwarf was unlikely to produce a detonation;
thus, such a scenario might produce a SN~Iax. However, \citet{Seitenzahl13} find that the conditions are met for a detonation in all of their models ranging from a single ignition point up to 1600 ignition points. Even a pure deflagration does not explain the incredibly low kinetic energy
for the most extreme members of the SN~Iax class, such as
SN~2008ha. 

To explore low-energy complete deflagration scenarios for more typical
SNe~Iax, we examined a homologous expansion model using the density
profiles of \citet{Ropke05}.  We find that it is possible to produce
the high densities observed at late times, but the initial kinetic
energy distribution must be several times lower than in the model
(also assuming the emitting region moves to lower velocities and
roughly constant density, as Figure \ref{fig:density}
requires). Scaling down the velocity of the density distribution, we
find that the speed of material at the radius with 10\% of the total
enclosed mass ($R_{0.1}$) must be as low as 150$\kms$ to approach the
observations.  At first glance, this result seems like it could
qualitatively explain the lack of the [\ion{O}{1}] $\lambda6300$ flux
by keeping a substantial amount of unburned material at high
density. However, quantitatively, we find that even for homologous
expansion that is scaled to the lower kinetic energies, at 600~days
after maximum brightness, \about 55\% of the oxygen should still be
below the critical density. This corresponds to \about 0.2 $M_\odot$
of low-density oxygen in the \citet{Ropke05} models, and is ruled out
by our observations at high significance (Figure \ref{fig:late_o}). 
However, our constraint could be weakened if the [\ion{O}{1}] excitation is being suppressed by a mechanism where photospheric lines are preferentially excited (\citealt{Sahu08}; see Section \ref{sec:oxygen}), though it is unclear whether this could continue as late as 600 days past maximum light.

Generically, we find that even if there is a photosphere hiding the
innermost regions, the models predict oxygen at higher velocities that
should be visible, and yet is not seen. Observationally, the
decreasing widths of the forbidden lines of iron and calcium with time
(see Figure~\ref{fig:velocity}) imply that these lines are not being
excited everywhere in the ejecta. If these lines are only produced
near the ``photosphere'' responsible for the numerous permitted Fe
lines, there must be a large reservoir of lower-density material at
higher velocities that is not radiating \emph{either} in these lines,
nor is it seen in lower density tracers like [\ion{O}{1}] $\lambda$6300.

\begin{figure*}[!t]
\includegraphics[width=0.8\textwidth]{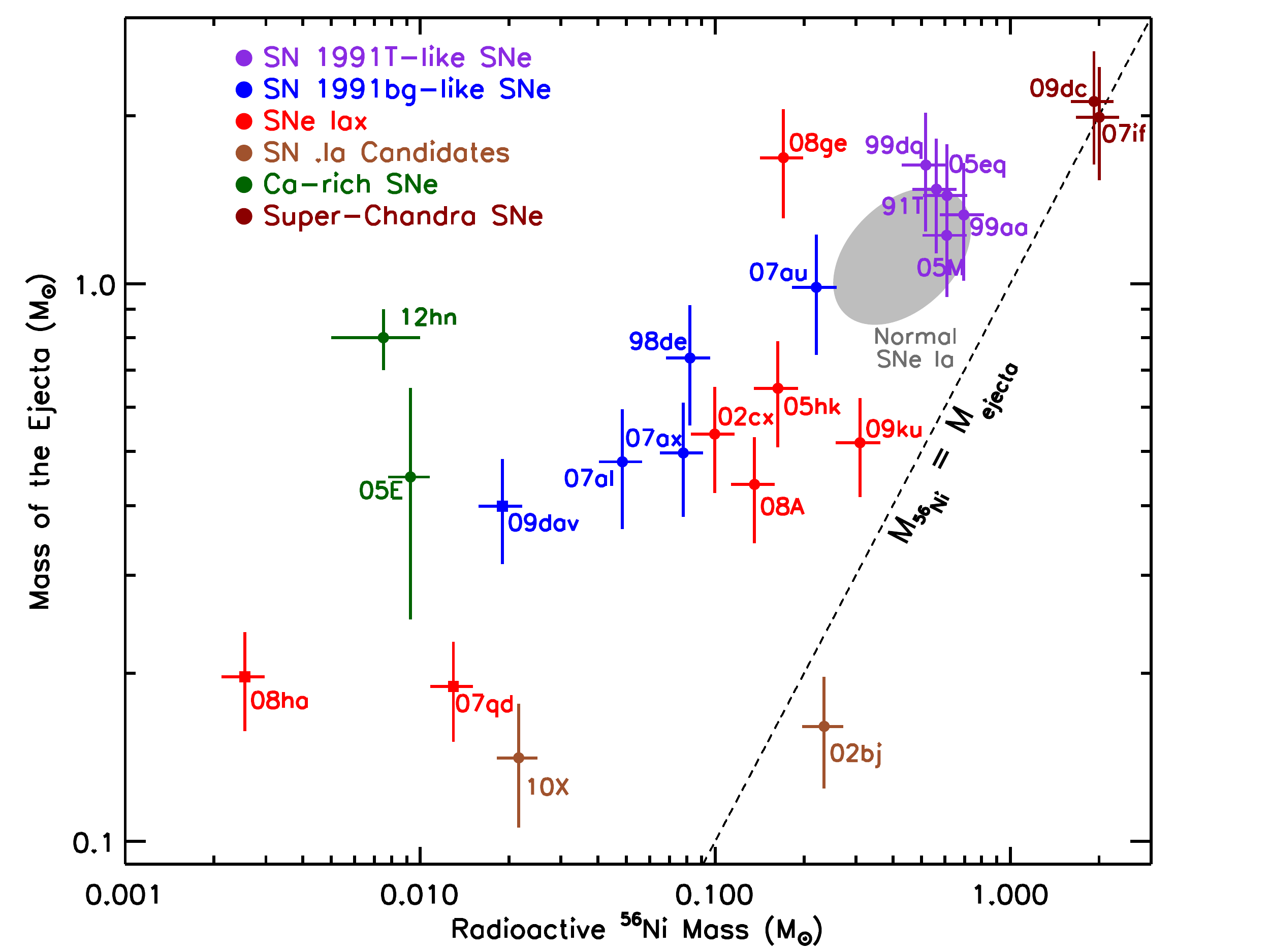}
\centering
\caption{Total ejected mass vs.\ radioactive nickel mass of
  white-dwarf SNe. Subclasses of objects are grouped by color as
  indicated in the plot. Circles denote typical members of the
  subclass, while squares show more extreme members. The dashed line
  denotes where the ejected mass equals the radioactive nickel
  mass. Objects to the right of the line (like SN~2002bj) require
  additional energy sources (beyond $^{56}$Ni radioactive power) to
  explain their luminosity.  References: SN~1991T
  \citep{Filippenko92T,Phillips92T}; SN~1998de, SN~2007al, SN~2007au, and PTF
  2009dav \citep{Sullivan11}; SN~1999aa and SN~1999dq
  \citep{Jha06_cfa2}; SN~2002bj \citep{Poznanski10}; SN~2002cx
  \citep{Li03,Jha06}; SN~2003du \citep{Hicken09}; SN~2003hv
  \citep{Leloudas09}; SN~2005E \citep{Perets10,Waldman11}; SN~2012hn \citep{Valenti14}; SN~2005M and
  SN~2005eq \citep{Ganeshalingam10}; SN~2005hk \citep{Phillips07};
  SN~2007ax \citep{Kasliwal08}; SN~2007if \citep{Scalzo10}; SN~2007qd
  \citep{McClelland10}; SN~2008A (this work); SN~2008ge
  \citep{Foley10_ge}; SN~2008ha \citep{Foley09,Valenti09}; SN~2009dc
  \citep{Taubenberger11}; SN~2009ku \citep{Narayan11}; and SN~2010X
  \citep{Kasliwal10}. Ancillary data were used in the analysis,
  including ejecta velocities \citep{Foley11}, rise times
  \citep{Ganeshalingam11}, bolometric corrections \citep{Contardo00},
  and distances and luminosities \citep{Ganeshalingam13}.}
\label{fig:ejecta_nickel}
\end{figure*}

In normal SNe~Ia, the widths of nebular lines at late times is
of order $10^{4}\kms$ \citep{Mazzali98,Silverman13}. If, as
expected, we are ``seeing through'' all the excited ejecta in these
forbidden transitions, the line profiles directly reveal the velocity
structure, and the observed line width is effectively the ``final''
velocity for the ejecta, $v_\infty$. Moreover, for an exploding white
dwarf that is completely disrupted, we expect $v_\infty \gtrsim v_{\rm
  esc}$, the escape velocity of that material in the white dwarf;
otherwise the explosion would need to be finely tuned to barely unbind
the white dwarf and leave the ejecta expanding at low velocities.

The widths of forbidden lines of \ion{Ca}{2} and \ion{Fe}{2} in
the late-time spectra of SN~2005hk are unprecedentedly low ---
500$\kms$. If this is interpreted as the velocity at infinity for the
emitting ejecta, it is an order of magnitude below the escape
velocity from the surface of the white dwarf, $v_\infty \ll v_{\rm
  esc}$. Using a simple model of a Chandrasekhar-mass white dwarf with
a polytropic equation of state, we calculated the escape velocity as a
function of the enclosed mass throughout the white dwarf. At $v_{\rm
  esc} = 500\kms$ the enclosed mass is just \about $5 \times 10^{-6}$
of the total mass. It seems implausible that all of the late-time
emission arises from just this innermost tiny fraction of the white dwarf.

More likely, we are witnessing a weak explosion, in which the
explosion energy is close to or less than the binding energy of the
white dwarf. The material emitting in the forbidden lines was barely
unbound, and we observe it at low $v_\infty \ll v_{\rm esc}$. In that
case, unless again the explosion energy is finely tuned, we would also
expect a significant fraction of the original white dwarf to not reach
escape velocity, and remain bound, leaving behind a remnant.  The high
densities observed might also be explained if the white dwarf was not
fully disrupted, with the mixed composition of the unbound ejecta
arising from a deflagration. \citet{Livne05} show that a single
ignition that is offset from the center of the white dwarf could form
a bubble that would convectively rise and break
through the surface, without fusion outside of the bubble. Events like
this could explain typical SNe~Iax and the extreme SN~2008ha, yet
still show the thermonuclear signature of \ion{Si}{2} and
\ion{S}{2}.

\citet{Kromer13} and \citet{Jordan12} find that a failed deflagration
of a white dwarf can produce properties similar to those observed for
SNe~2005hk and 2008A. A key prediction of these models is that the
explosion leaves behind a bound remnant. This is consistent with our
argument above about the escape velocity of the white
dwarf. \citet{Jordan12} find that the remnant receives a kick of
\about $500 \kms$, which is roughly consistent with our measurements
of the velocity offsets from the host galaxies. This assumes that the
velocities of the forbidden lines are measuring the bulk motion of the
ejecta rather than an excitation effect. While \citet{Kromer13} do not
find such significant kick velocities, their model matches the
bolometric light curves of SN~2005hk and SN~2008A relatively well,
except at the latest observed epochs (Figure
\ref{fig:bolometric}).

While suggestive, the case for a bound remnant is not without
problems. If we assume homologous expansion, material at 300$\kms$,
about half the width of the forbidden lines, would be at 100~AU about
600~days past explosion. This cannot be the radius of a true
photosphere, because the bolometric luminosity at that epoch (\about
$10^{38}$~erg~s$^{-1}$) is much too low given the estimated
temperature. The bolometric luminosity limits the photospheric radius
to $\lesssim 3.5$ AU, making it difficult to directly connect the high
density expanding material with a potential bound remnant. Moreover,
these models still require an explanation for the lack of unburned
material detected in ejecta that should be well mixed (including
intermediate-mass and iron-group elements) and reach low
density. Perhaps additional complexity is required, such as back-warming 
in which inwardly traveling gamma-rays deposit their energy and heat
lower-velocity regions, while outwardly traveling gamma-rays escape.

Throughout this paper, we have assumed that radioactive heating from
$^{56}$Ni dominates the luminosity of these SNe~Iax. It is instructive
to put SNe~Iax in context with other probable thermonuclear explosions
of white dwarfs. We compiled a sample of such objects from the
literature to infer their radioactive nickel mass and ejecta mass; the
results are shown in Figure \ref{fig:ejecta_nickel}. The $^{56}$Ni
mass was calculated using the \citet{Arnett82} rule, and for this
comparison we simply adopted a uniform uncertainty of 15\% for all
bolometric luminosities and a 10\% uncertainty for the SN rise
times. These correspond to typical published uncertainties for the
objects for which error bars were reported. We calculated the ejecta
mass following \citet{Foley09}, assuming an opacity of $\kappa = 0.1$~cm$^2$~g$^-1$ and a uniform 5\%
uncertainty on the ejecta velocities. The relations are scaled such
that normal SNe~Ia produce 1.4 $M_\odot$ of ejecta, with \about 0.5
$M_\odot$ of $^{56}$Ni. Normal SNe~Ia form an approximate continuum
from the subluminous SN 1991bg-like objects
\citep{Filippenko92bg,Leibundgut93} to the superluminous SN 1991T-like
\citep{Filippenko92T,Phillips92T}, and even to more extreme
``super-Chandra'' objects such as SN~2009dc
\citep{Howell06,Scalzo10,Taubenberger11,Silverman11}. 

The three SNe~Iax studied in detail in this work cluster below the
SN~1991bg-like objects with \about 0.5 $M_{\odot}$ of ejecta and 0.15
$M_{\odot}$ of radioactive nickel. However, the SNe~Iax class shows
significant diversity in both axes: SN~2009ku is very close to having
burned a large fraction of its ejecta into $^{56}$Ni, while SN~2008ge
shows lower amounts of radioactive nickel than normal SNe~Ia, but a
similar ejecta mass. Two of the most extreme members of the SNe~Iax
subclass are SN~2008ha and SN~2007qd \citep{McClelland10}; both events
produced low ejecta mass and $^{56}$Ni mass, similar to other
low-luminosity transients like the ``calcium-rich'' SN~2005E
\citep{Perets10} or the ``SN~.Ia'' candidate SN~2010X
\citep{Bildsten07,Kasliwal10}. Taken together, the luminosities, rise
times, and ejecta velocities of SNe~Iax certainly do not contradict
the hypothesis that SNe~Iax are thermonuclear explosions of white
dwarfs in which radioactive nickel provides the UVOIR light, though
the observations do pose a significant challenge to models. We note that \citet{Sahu08} were able to fit the bolometric light curve and some of the spectral features of SN~2005hk with an explosion model that assumed an ejecta mass of 1.4 $M_\odot$ which implies that more complex modeling may be necessary.

\section{Summary}

We present ground-based and {\it HST} photometry and spectroscopy of
SNe~2005hk and 2008A, two typical SNe~Iax. These objects remain at
high density for \about 2 yr after explosion and are not observed
to enter the typical nebular phase up to 400~days after $B$-band
maximum brightness. We find no evidence for unburned material at low
velocities, either directly through spectroscopy or indirectly with our
{\it HST} photometry. Based on emission-line diagnostics, we find that
the density of the emitting region remains roughly constant over the
duration of the observations, though the widths of even the forbidden
lines decrease. We do not see the signature of the IR
catastrophe in optical colors. The bolometric luminosity of SN~2005hk
and SN~2008A fades more slowly than that of normal SNe~Ia at 100--200~days
after maximum brightness, but then declines faster than in normal SNe~Ia
at phases of 400--600~days. Failed deflagration models that leave a
bound remnant \citep[e.g.,][]{Jordan12, Kromer13} show promise for
explaining these explosions, but no single proposed model can explain all 
of our observations.

\acknowledgments

We dedicate this paper to the lasting memory of our dear friend and
cherished colleague, Weidong Li.

We thank Mark~Phillips and Roger~Romani for help in acquiring the {\it
  HST} observations of SN~2005hk. We acknowledge usage of the NASA
Extragalactic Database (http://ned.ipac.caltech.edu) and the HyperLeda
database (http://leda.univ-lyon1.fr).

This research at Rutgers University was supported through NASA/{\it
  HST} grants GO-11133.01 and GO-11590.01, along with U.S. Department
of Energy (DOE) grant DE-FG02-08ER41562, and National Science
Foundation (NSF) CAREER award AST-0847157 to S.W.J., and a GAANN
Fellowship to C.M.  J.M.S. is supported by an NSF Astronomy and
Astrophysics Postdoctoral Fellowship under award AST-1302771. The
research of J.C.W. is supported in part by NSF Grant AST-1109801.
G.L. is supported by the Swedish Research Council through grant
No. 623-2011-7117. A.V.F. and his group at UC Berkeley are funded by
Gary and Cynthia Bengier, the Richard and Rhoda Goldman Fund, NSF
grant AST-1211916, the TABASGO Foundation, and NASA/{\it HST} grants
GO-10877 and AR-12623.

Support for \textit{HST} programs GO-10877, GO-11133, GO-11590, and AR-12623
was provided by NASA through a grant from the Space Telescope Science
Institute, which is operated by the Association of Universities for
Research in Astronomy, Incorporated, under NASA contract NAS5-26555.

Some of the data presented herein were obtained at the W. M. Keck
Observatory, which is operated as a scientific partnership among the
University of California, the California Institute of Technology, and
NASA, made possible by the generous financial support of the
W. M. Keck Foundation. The authors recognize and acknowledge the very
significant cultural role and reverence that the summit of Mauna Kea
has always had within the indigenous Hawaiian community, and we are
most privileged to have the opportunity to explore the Universe from
this mountain.

Funding for the SDSS and SDSS-II has been provided by the Alfred
P. Sloan Foundation, the Participating Institutions, the NSF, the 
U.S. Department of Energy, NASA, the Japanese Monbukagakusho, the
Max Planck Society, and the Higher Education Funding Council for
England. The SDSS Web site is http://www.sdss.org/.
The SDSS is managed by the Astrophysical Research Consortium for the
Participating Institutions. The Participating Institutions are the
American Museum of Natural History, Astrophysical Institute Potsdam,
University of Basel, University of Cambridge, Case Western Reserve
University, University of Chicago, Drexel University, Fermilab, the
Institute for Advanced Study, the Japan Participation Group, Johns
Hopkins University, the Joint Institute for Nuclear Astrophysics, the
Kavli Institute for Particle Astrophysics and Cosmology, the Korean
Scientist Group, the Chinese Academy of Sciences (LAMOST), Los Alamos
National Laboratory, the Max-Planck-Institute for Astronomy (MPIA),
the Max-Planck-Institute for Astrophysics (MPA), New Mexico State
University, Ohio State University, University of Pittsburgh,
University of Portsmouth, Princeton University, the United States
Naval Observatory, and the University of Washington.

\clearpage 

\bibliography{sn2008a}        

\end{document}